\documentclass{jfm}
\usepackage{graphicx}
\usepackage{mathtools}
\usepackage{epstopdf, epsfig}
\usepackage{tikz}
\usepackage[ruled]{algorithm2e}
\usepackage{booktabs}
\usepackage{array}
\usepackage{amsmath}
\usepackage{multirow}
\usepackage{comment}
\usepackage{natbib}
\usepackage[]{color}
\usepackage[]{xcolor}
\usepackage{todonotes}
\usepackage{placeins}
\usepackage{relsize}
\usepackage{soul}

\newcolumntype{x}[1]{>{\centering\arraybackslash\hspace{0pt}}p{#1}}

\newcommand\norm[1]{\left\lVert#1\right\rVert}
\newcommand{\solution}{\mathbf{q}}
\newcommand{\transpose}{\mathrm{T}}
\newcommand{\cntrVec}{\mathbf{c}}

\newcommand{\cost}{\mathcal{J}}
\newcommand{\meas}{\mathbf{m}}
\newcommand{\projCntrVec}{\mathcal{M}(\solution)}
\newcommand{\projCntrVecMean}{\mathcal{M}(\solution_i)}
\newcommand\projCntrVecEns[1]{\mathcal{M}(\solution^{\scriptscriptstyle{(#1)}})}
\newcommand{\covCntrVec}{\Sigma_\mathbf{c}}
\newcommand{\covMeasVec}{\Sigma_\mathbf{m}}
\newcommand{\Weight}{\mathbf{W}_{\!s}}
\newcommand{\pertMat}{\mathbf{P}}
\newcommand{\obsMat}{\mathbf{H}}
\newcommand{\hessian}{\mathcal{H}}
\newcommand{\gradLogLiklihood}{\mathbf{g}}
\newcommand{\weights}{\mathbf{w}}
\newcommand{\Nens}{{N}}
\newcommand{\expectation}{\mathbb{E}}
\newcommand{\identity}{\mathbf{I}}
\newcommand{\AMat}{\mathbf{A}}
\newcommand{\BMat}{\mathbf{B}}
\newcommand{\imag}{\mathrm{i}}

\newcommand{\measHorizon}{T_{a}}
\newcommand{\measFrequency}{\Delta t_a}

\newcommand{\LinCost}{\widetilde{\cost}}

\newcommand{\mode}[2]{({#1},{#2})}
\newcommand{\modeFkz}[2]{(F,k_z)=({#1},{#2})}

\allowdisplaybreaks[1]

\shorttitle{Optimal interpretation of wall observations}
\shortauthor{D. A. Buchta and T. A. Zaki}
\title{Observation-infused simulations of high-speed boundary layer transition}

\author{David A. Buchta\aff{1}
 \and Tamer A. Zaki\aff{1}   \corresp{\email{t.zaki@jhu.edu}}}

\affiliation{\aff{1}Department of Mechanical Engineering, Johns Hopkins University,
Baltimore, MD 21218, USA}
\begin{document}

\maketitle

\begin{abstract}
High-speed boundary-layer transition is extremely sensitive to the free-stream disturbances which are often uncertain. This uncertainty compromises predictions of models and simulations. To enhance the fidelity of simulations, we directly infuse them with available observations. Our methodology is general and can be adopted with any simulation tool, and is herein demonstrated using direct numerical simulations. An ensemble variational (EnVar) optimization is performed, whereby we determine the upstream flow that optimally reproduces the observations. The cost functional accounts for our relative confidence in the model and the observations, and judicious choice of the ensemble members improves convergence and reduces the prediction uncertainty. We demonstrate our observation-infused predictions for boundary-layer transition at Mach 4.5. Without prior knowledge of the free-stream condition, and using only observations of wall pressure at isolated locations from an independent computation (true flow), all of the relevant inflow disturbances are identified. We then evaluate the entire flow field, beyond the original limited wall observations, and interpret simulations consistently with data and vice versa. Our predicted flow compares favorably to the true `unknown' state, and discrepancies are analyzed in detail. We also examine the impact of weighting of observations. Improved convergence of the inverse problem and accuracy of the inflow amplitudes and phases are demonstrated, and are explained by aid of a simple example from two-dimensional unstable, chaotic convection.
\end{abstract}

\section{Introduction}\label{sec:Introduction}

Accurate interpretation of wall measurements in high-speed transitional boundary layers is challenging due to sensor limitations, scarcity of data, and the amplification of uncertainties\textemdash the last two factors being the focus of the present work. The acquired finite-amplitude wall signal is much smaller in dimension than the flow state; And it is the outcome of antecedent events that are governed by the Navier--Stokes equations which are chaotic and hence very sensitive to uncertainties. In order to optimize the interpretation of transition data, we develop an ensemble variational (EnVar) technique that infuses direct numerical simulations (DNS) with available observations, here wall pressure data at discrete sensor locations.  The data is obtained from an independent computation of $M=4.5$ transitioning flat-plate boundary layer (see figure~\ref{fig:configuration}~(a)). Using our EnVar approach, we (i) identify the amplitudes and phases of unknown inflow instabilities that (ii) reproduce observations, (iii) reconstruct the full flow field beyond the spatio-temporal limits of the wall data, and (iv) decode the transition mechanism. Since the location of the wall-pressure probes encode varying levels of flow information and sensitivity, sensor weightings are considered to enhance interpretation. We provide a clear mathematical interpretation of which sensors should be discounted in order to improve accuracy.

Advances in DNS have enabled precise descriptions of hypersonic boundary-layer transition in progressively more complex regimes.  For example, transition was examined on sharp~\citep{sivasubramanian2015direct}, flared~\citep{hader2019direct}, and blunted~\citep{balakumar2018transition} cones, as well as the flow over more complex geometries such as the BOLT design~\citep{knutson2019numerical} including the effects of micron-sized wall roughness elements~\citep{thome2018direct}. Simulation data provide the complete space-time solution of the Navier--Stokes equations; transition mechanisms can then be directly probed and clarified. Recently, for Mach 4.5 flow, \cite{jahanbakhshi_zaki_2019} computed the energy transfer among wavenumber triads, as done in \cite{cheung2010linear}, to detail the mode interactions that lead to the earliest possible transition on an adiabatic flat plate. Tightly controlled DNS also provides a framework for identifying plausible hypersonic transition mechanisms~\citep{franko2013breakdown,franko2014effect,hader2019direct} and discerning valid regimes of parallel linear stability theory, linear parabolized stability equations~(PSE) and their nonlinear extension~\citep{chang1993linear,zhao2016improved}. 

Common among all simulations are assumed initial-boundary conditions, which exposes the uncertainty of how `real' environmental conditions may alter transition. Often, mean inflows are based on similarity profiles of the Blasius form; yet, upstream shocks can distort the base flow and couple to boundary-layer disturbances influencing transition~\citep{chang1990effects,pinna2013effects}. To consider the shock and leading edge effects, precursor simulations over the entire test article, often of lower fidelity, provide the post-shock inflow state to truncated DNS domains~\citep{li2010direct,laible2011numerical,sivasubramanian2015direct}.  Atop the base flow, near or at the inflow plane,  space-time dependent disturbances are then introduced to seed transition.  Excitation is usually a superposition of waves in one or more flow quantities: free-stream pressure~\citep{balakumar2018transition,hader2018towards}, wall-normal velocity perturbations~\citep{li2010direct,laible2011numerical,franko2014effect,sivasubramanian2015direct,hader2019direct}, or momentum sources~\citep{knutson2019numerical}. And, the superposition is often limited to a handful of wavenumber-frequencies. Broadband effects have also been introduced, implicitly, by pseudo random fluctuations~\citep{li2010direct,knutson2019numerical,hader2018towards}. Similarly, transition can be triggered directly by imposing unstable mode shapes comprised of vortical, acoustic, and thermal modes~\citep{pruett1998direct}. While these studies expose numerous routes to turbulence, none has been directly constrained to match independent observations.

When theory or simulation conditions are incongruous with experiments, deviations are inevitable. Discrepancy often leads to speculation and errors being ascribed to uncertainty in environmental or surface conditions.
\citet{lysenko1984effect} explained the deviations in measured and theoretical instability growth rates in cooled flows, in part, in terms of the angle of waves infiltrating the boundary layer. \cite{park_zaki_2019} performed adjoint sensitivity analyses that detail how uncertainty in mean-velocity and temperature profiles affect growth rates; they showed that the measured growth rates by \cite{lysenko1984effect} can result from 1-2\% uncertainty in wall temperature. 

In another example relevant to uncertainty of inflow, DNS of $M=6$ flow over a flared cone considered randomized inflow pressure fluctuations, in place of the free-stream disturbance environment of the tunnel~\citep{hader2018towards}. Simulations revealed agreement with the number of azimuthal high-temperature streaks to experiments as well as similar streamwise temperature variation.  Together with other tightly controlled `breakdown' simulations~\citep{hader2019direct}, the DNS provided a clear interpretation of the streak mechanism. However, the wall pressure fluctuations for controlled breakdown DNS were over-predicted, up to 15$\times$~\citep{Chynoweth2019history}. The forcing amplitude was selected to ensure \textit{that the nonlinear breakdown would occur in the computational domain}~\citep{hader2018towards} rather than reproducing the experimental environment.  While the basic transition mechanism appears to be explained by fundamental breakdown, i.e.\,interacting two- and three-dimensional waves at identical second-mode frequencies, the simulations do not recover the precise experimental trajectories.  Reproducing experimental measurements from \textit{all} available sensor modalities would cement trust in the interpretation of data and in the computational model. 

In addition to advancing direct simulations, experiments are gaining unprecedented access to hypersonic boundary layer flow,  as well as control of the flow environment and test article design (e.g.,~leading edge bluntness and surface roughness). Although flight tests are the ultimate target for achieving realism, flow parameters in flight (e.g.,~Mach and Reynolds numbers) are changing simultaneously, together with an uncertain environment, which confound predictions and interpretation of data~\citep{thiele2018instrumentation}. Ground-based tests, however, offer precise control of the nominal Mach and Reynolds numbers, thereby isolating the effects of uncertain free-stream disturbances impinging on the boundary layers, as this remains a leading uncertainty in decoding transition mechanisms~\citep{reshotko1976boundary,bushnell1990notes,saric2002boundary,schneider2015developing}, in part, because they are difficult to fully characterize. 

Despite a looming presence of an uncertain disturbance environment, state-of-the art boundary-layer measurements illuminate hypersonic transition mechanisms. Piezoelectric pressure sensors (e.g.\,PCB-132) provide time resolution to detect high-frequency instabilities, their harmonics, and subsequent spectral broadening during transition. \cite{marineau2019analysis} provide a compilation of surface pressure measurements taken in 11 wind tunnels for $5\leq M\leq 14$ and unit Reynolds number $1.5 \leq Re/m \leq 16 $ million. In addition to PCB wall pressure identifying mode amplification rates, spatially-resolved temperature measurements, using temperature sensitive paint (TSP), reveal striking, closely packed, streamwise streaks which intensify, cool, and then reheat downstream on straight and flared cones in Mach 6 flow~\citep{chynoweth2018measurements,Chynoweth2019history}. Additionally, time-resolved schlieren measurements~\citep{laurence2012time,laurence2016experimental,kennedy2018investigation} have quantified the structure, spectral properties, and advection speed of the rope-like second-mode instabilities and their breakdown to turbulence~\citep{casper2016hypersonic}.  On more complex geometries such as BOLT, infrared cameras, PCB sensors, and TSP probe the transition location and the structure of hot and cold temperature streaks in the presence of surface roughness~\citep{berridge2019hypersonic} and variations of free-stream disturbances found in different wind tunnels~\citep{kostak2019freestream}.

Still, similar to simulations, the free-stream disturbances in experiments are a key source of uncertainty~\citep{schneider2001effects}. And, when pointwise free-stream measurements are taken, for example by flow intrusive PCB enabled Pitot tubes~\citep{bounitch2011improved,masutti2012disturbance}, complex flow interactions by preceding bow shock and concomitant stagnation region confound interpretation of the signal, which has further spurred detailed modeling efforts using simulation~\citep{chaudhry2017computing,Duan2019characterization}. Recently, \citet{parziale2014free}~have quantified free-stream density fluctuations using focused laser differential interferometer (FLDI), which increases temporal and spatial resolution relative to pressure probes. Paramount, FLDI is a non-intrusive optimal technique making interpretation of data straightforward relative to their flow-intrusive counterparts, and they, like schlieren, have also been successful in characterizing wave packets associated with second-mode boundary-layer instability~\citep{parziale2015observations}.


The hypersonics community has recognized the potential of interfacing experimental data with computations~\citep{schneider2015developing,Chynoweth2019history,Duan2019characterization}. For example, comparisons with PSE show success for predicting instability amplification and flow structure for certain regions of measurement before transition to turbulence~\citep{parziale2015observations,kennedy2018investigation,marineau2019analysis}.  Also, \cite{Duan2019characterization} considered tunnel-scale simulations of the Mach 6 Hypersonic Ludwieg Tube, Braunschweig (HLB), which capture free-stream disturbances originating from turbulent boundary layers entering the test section; Simulations were used in tandem with free-stream point-wise measurements to characterize the spectrum, direction of incident waves, and transfer function between physical waves and acquired signals. However, embedding this realism into transition models is still an active area of research. In addition, extension to hypersonic flight may be intractable, as of now, because it requires accurate precursor simulations of the flight environment in the troposphere and stratosphere. Recognizing the difficulty to fully characterize the free stream, \cite{schneider2015developing} reflects that direct simulations will need calibrated models of incoming disturbance fields from a combination of simulations and free-stream measurements.

Absent a complete characterization of the free-stream condition, the present work attempts to solve the inverse problem: Starting with available downstream observations, here wall-pressure data, we reconstruct the unknown inflow disturbance for simulations of flat-plate boundary-layer transition. The approach exploits the fact that observation-data are encoded with antecedent events in the flow, which when decoded can be traced back to the inflow disturbance. Once the unknown inflow is discovered, the full flow field is also available, far surpassing the domain extents and resolution of the original observations.  
Therefore, the approach is attractive for interpretation of flight data, e.g.\,reconstructing the flight environment from on-board measurements. 

Section~\ref{sec:DAFormulation} introduces the ensemble-variational (EnVar) framework for observation-infused simulations. The first EnVar step approximates the inflow using a linear PSE model.  In this limit, the framework simplifies to a direct inversion to acquire the unknown inflow modes. The initial interpretation is further improved using EnVar iterations with DNS to extend its predictive capacity into the nonlinear regime. The results are given in \S\ref{sec:results}, and \S\ref{sec:error} analyzes the mechanisms for deviation between the truth and the optimally interpreted inflow. 
Section~\ref{sec:optimalWeighting} analyzes the effect of observation weighting since (i) not all sensors may yield useful inflow information and (ii) EnVar optimization may be peculiarly sensitive to sensors recording chaotic fluctuations such as those in turbulence.  Finally, discussion and conclusions are provided in \S\ref{sec:summary}. Analysis of EnVar performance are provided in Appendices~\ref{app:accuracy} and \ref{app:lorenz} for boundary layer transition, and unstable thermal convection, respectively.

\section{Formulation of EnVar observation-infused simulations}\label{sec:DAFormulation}

\begin{figure}
        \centering
        \includegraphics[width=\textwidth]{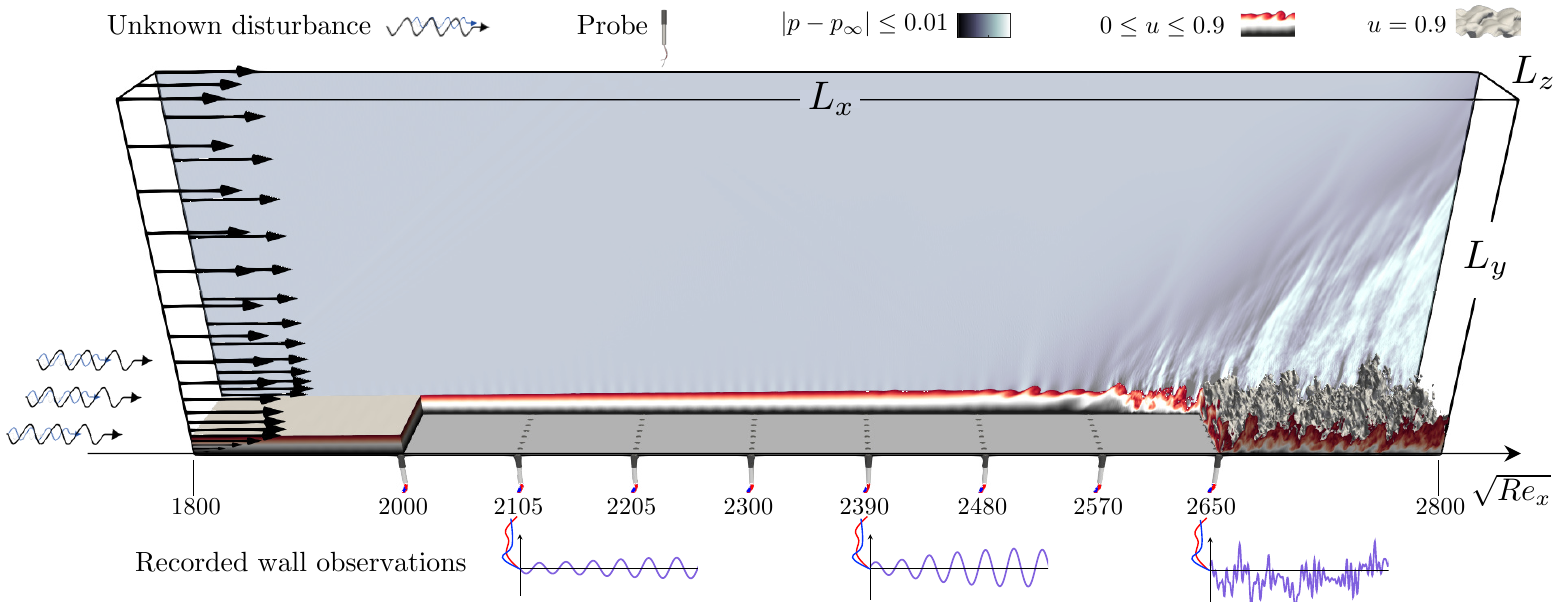}
        \caption{Flow configuration for a transitional high-speed boundary layer on a flat plate. Cutout shows the locations of discrete wall-pressure probes.}
        \label{fig:configuration}
\end{figure}

In order to demonstrate the framework for observation-infused simulation, we adopt a canonical flat-plate boundary layer as a test-bed (see figure \ref{fig:configuration}). 
This choice enables us to relate the formulation to a physical example.
However, the formulation is general, applicable to any flow configuration, available observations, and numerical simulation method.
Where there are elements specific to the flat-plate configuration, they are indicated as such.

We assume access to limited measurements $\meas$, or observations, which will be taken to be a finite number of wall-pressure traces.
We attempt to predict unknown parameters of the simulations $\cntrVec$, which will be related to the inflow disturbances.
With an estimate of $\cntrVec$, we can simulate the flow using the three-dimensional Navier-Stokes equations for a calorically perfect gas, formulated in Cartesian coordinates,  
\begin{align}\label{eq:mass}
    \frac{\partial \rho}{\partial t} + \frac{\partial (\rho u_i)}{\partial x_i} &= 0, \\\label{eq:mom}
    \frac{\partial (\rho u_i)}{\partial t} + \frac{\partial }{\partial x_j}(\rho u_i u_j) &= -\frac{\partial p}{\partial x_i} + \frac{\partial \tau_{ij}}{\partial x_j}, \\\label{eq:energy}
    \frac{\partial E}{\partial t} + \frac{\partial }{\partial x_i}[u_i(E+p)] &= \frac{\partial}{\partial x_i} (\tau_{ij} u_j) + \frac{\partial}{\partial x_i} \!\left(\kappa \frac{\partial T} {\partial x_i}\right),
\end{align}
where $\rho$ is the density, $u_i$ is the velocity in the $i$-th coordinate direction, $p$ is the pressure, $E=p/(\gamma-1) + 1/2\rho u_i u_i$ is the total energy, $\kappa$ is the thermal conductivity, and $T=p/(\rho R)$ is the temperature with gas constant $R$ and ratio of specific heats $\gamma$. The components of the viscous stress tensor, assuming a Newtonian fluid, is
\begin{equation}
    \tau_{ij} = \mu \left(\frac{\partial u_i}{\partial x_j}+\frac{\partial u_j}{\partial x_i} \right) + \left(\mu_b - \frac{2}{3}\mu\right) \delta_{ij}\frac{\partial u_k}{\partial x_k},
\end{equation}
where $\mu$ and $\mu_b$ are the shear and bulk viscosity, respectively. In compact operator form, the solution state $\mathbf{q}=\left[\rho, \rho \mathbf{u}, E \right]^{\mathrm{T}}$ which solves (\ref{eq:mass}-\ref{eq:energy}) is denoted as $\mathbf{q}=\mathcal{N}(\mathbf{c})$. For a linear model, e.g.~PSE, we will adopt the notation $\mathcal{L}(\mathbf{c})$. 
From the flow solution, we can collect model observations $\projCntrVec$ and compare them to $\meas$.

The optimal value of the unknown $\cntrVec$ can be identified by minimizing the cost function, 
\begin{equation}
\cost = \frac{1}{2}\norm{\meas - \projCntrVec }^2_{\covMeasVec^{-1}} + \frac{1}{2}\norm{\cntrVec-\cntrVec_i}^2_{\covCntrVec^{-1}}.  
\label{eq:cost}
\end{equation}
The first term quantifies the deviation between observations and the model output; the second term serves to regularize the behavior of $\cost$ limiting how far subsequent control vectors can deviate from their initial value $(\mathbf{c}_i)$; the $\covMeasVec$ and $\covCntrVec$ are the covariance matrices for the observations and control vector, respectively; and the operator $\norm{\cdot}^2_{\Sigma^{-1}}$ is $(\cdot)^\transpose \Sigma^{-1} (\cdot)$.  The covariance matrices in (\ref{eq:cost}) serve to weight the relative trust between the data (first term) and the control vector (second term). Based on Bayes' theorem, the cost represents the negative logarithm of the posterior probability of identifying $\cntrVec$ given $\meas$; this function is comprised of the likelihood of matching $\meas$ given $\cntrVec$ balanced by the prior trust in $\cntrVec_i$. 

An efficient approach to minimize $\cost$ is to seek a control vector, iteratively, in the negative gradient direction, $\cntrVec_{i+1}=\cntrVec_i-\gamma \nabla\cost$ by a scalar step size $\gamma$. 
The evaluation of the gradient $\nabla\cost$ can be performed using various techniques, each with advantages and drawbacks.  Adjoint approaches, for example, provide accurate gradient and their computational cost is independent of the size of the control vector; however they require an adjoint model and storage of the forward solution, and accuracy of the forward-adjoint relation is difficult to maintain for long integration times~\citep{vishnampet2015practical,wang2019discrete,WangZaki_arxiv}. An alternative, which we adopt here, is an ensemble variational (EnVar) technique, e.g.~\citet{liu2008ensemble,mons2016reconstruction}, where an accurate approximation of the local gradient at $\cntrVec_{i}$ is evaluated from an ensemble of control vectors.

In EnVar, updates to the control vector $\cntrVec_i$ in the gradient direction are made by a weighted ($\weights$) superposition,
\begin{equation}
\cntrVec_{i+1} = \cntrVec_{i} +  \pertMat \weights,
\label{eq:superposition}
\end{equation}
where $\pertMat=\left[\cntrVec^{\scriptscriptstyle{(1)}}-\cntrVec_{i} \,\,|\,\, \cntrVec^{\scriptscriptstyle{(2)}}-\cntrVec_{i} \,\,|\,\, \ldots \,\,|\,\, \cntrVec^{\scriptscriptstyle{(\Nens)}}-\cntrVec_{i} \right]$ is a matrix with each column equal to the deviation of an ensemble member $\cntrVec^{\scriptscriptstyle{(j)}}$ from the mean $\cntrVec_{i}$, and $\cntrVec^{\scriptscriptstyle{(j)}}$ are sampled from a Gaussian distribution around that mean and with covariance~$\covCntrVec$. Note that in (\ref{eq:superposition}) the control vector is now effectively $\weights$, and the gradient and step-size are encoded in the matrix-vector multiplication $\pertMat \weights$, which will be discussed in detail.  

Assuming that the ensemble members are small perturbations to the mean control vector, a linear approximation is valid and the model observations are expanded as,
\begin{equation}
\projCntrVec \approx 
    \projCntrVecMean + \frac{\partial \projCntrVec}{\partial \cntrVec}\bigg{|}_{\cntrVec_i}(\cntrVec-\cntrVec_i) 
    = \projCntrVecMean + \underbrace{\frac{\partial \projCntrVec}{\partial \weights}\bigg{|}_{\cntrVec_i}}_{\obsMat} \underbrace{\frac{\partial \weights}{\partial \cntrVec}\bigg{|}_{\cntrVec_i}}_{\pertMat^{-1}} \pertMat \weights 
    = \projCntrVecMean + \obsMat \weights,
\label{eq:obsLinearization}
\end{equation}
where
\begin{equation}\nonumber
    \obsMat=\left[\projCntrVecEns{1}-\projCntrVecMean \,\,|\,\, \projCntrVecEns{2}-\projCntrVecMean \,\,|\,\, \ldots \,\,|\,\, \projCntrVecEns{\Nens}-\projCntrVecMean\right],
\end{equation}
is the observation matrix, with columns equal to the deviation of model observations from each ensemble member and from the mean. Substitution of (\ref{eq:superposition}) and (\ref{eq:obsLinearization}) in (\ref{eq:cost}) yields 
\begin{equation}
\LinCost = \frac{1}{2}\norm{\meas - \projCntrVecMean - \obsMat \weights}^2_{\covMeasVec^{-1}} + \frac{1}{2}\norm{\pertMat\weights}^2_{\covCntrVec^{-1}},
\label{eq:costLinearization}
\end{equation}
which is straightforward to differentiate with respect to $\weights$ to obtain the gradient and Hessian, 
\begin{align}
    {\partial \LinCost \over \partial \weights} &=   \obsMat^\transpose \covMeasVec^{-1} \left\{\projCntrVecMean  + \obsMat \weights - \meas\right\} + \pertMat^\transpose \covCntrVec^{-1} \pertMat \weights  
    \label{eq:GradJ} \\
    \hessian &\equiv {\partial^2 \LinCost \over \partial \weights^2} = \obsMat^\transpose \covMeasVec^{-1} \obsMat + \pertMat^\transpose \covCntrVec^{-1} \pertMat,
    \label{eq:HessianJ}
\end{align}
which can then be adopted to minimize the cost using a variety of methods, e.g.\,conjugate gradient~\citep{wang2019spatial}, L-BFGS~\citep{wang2019discrete}, \etc In the present study, we use a Newton step, where the gradient is set to zero and corresponding weights are acquired by
\begin{equation}
    \weights = -\bigg{(}\underbrace{\obsMat^\transpose \covMeasVec^{-1} \obsMat 
+ \pertMat^\transpose \covCntrVec^{-1} \pertMat }_{\hessian}\bigg{)}^{-1}
\bigg{(}\underbrace{\obsMat^\transpose \covMeasVec^{-1} \left\{\projCntrVecMean  - \meas\right\}}_{\gradLogLiklihood}  \bigg{)}.
\label{eq:weights}
\end{equation}
The success of reducing $\cost$ with the Newton step (\ref{eq:weights}) depends, in part, on obeying the tangent approximation, and appendix~\ref{app:accuracy} shows the effect of nonlinearity on an EnVar gradient approximation and viable step sizes in the gradient direction. 
The expression (\ref{eq:weights}), compactly expressed as $\weights=-\hessian^{-1} \gradLogLiklihood$, is made of two key quantities, where $\hessian$ is the Hessian matrix of $\LinCost$ and $\gradLogLiklihood=\nabla \LinCost(\weights=0)$ is the gradient at the current control vector. The first factor in (\ref{eq:weights}) has significant bearing on the difficulty, accuracy and robustness of the solution to the inverse problem. Part of the the second factor indicates that the update to the previous control vector in (\ref{eq:superposition}) is proportional to the deviation between the model predictions for the mean control vector and observations. The iterative formula (\ref{eq:superposition}) continues until the cost functional reduces to a prescribed percentage of its initial value or until the change in cost stagnates. 

%
Covariance matrices in (\ref{eq:costLinearization}) model the relative uncertainty, or lack of trust in the observation data and in the prior control vector.  The former, $\covMeasVec$, is based on the acquisition error from diagnostic equipment, for example irreducible signal noise or the systematic error in the  mathematical model to convert an electrical signal to a flow quantity. The $\covCntrVec$ reflects lack of confidence in the prior mean control vector. During an iterative optimization scheme, as $\cost$ is progressively reduced, the trust in $\cntrVec$ should increase; thus the $\covCntrVec$ should be reduced. In practice, at the end of each iteration, the covariance of the ensemble members is updated by taking advantage of the local shape of the cost function.  Specifically, ensemble-member perturbations are generated taking into account the Hessian of the cost function, 
\begin{equation}
    {\pertMat_{\scriptscriptstyle{{i+1}}} = \sqrt{\Nens-1}\pertMat_{\scriptscriptstyle{{i}}} \hessian^{-1/2} \mathbf{U}}
    \label{eq:pertMatPosterior}
\end{equation}
where $\mathbf{U}$ is a random, mean preserving ($\mathbf{U} \mathbf{1} = \mathbf{1} $), unitary ($\mathbf{U}^\transpose \mathbf{U} = \mathbf{I}$) matrix.  The covariance for the $\covCntrVec$ is then updated to reflect these statistics, 
\begin{equation}
    \covCntrVec^{\scriptscriptstyle{{i+1}}}=\frac{1}{\Nens-1}\pertMat_{\scriptscriptstyle{{i+1}}} {\pertMat_{\scriptscriptstyle{{i+1}}}}^{\!\!\!\!\!\!\!\transpose}.
    \label{eq:covariancePosterior}
\end{equation}
A derivation of (\ref{eq:pertMatPosterior}-\ref{eq:covariancePosterior}) is provided in Appendix~\ref{app:matrixIdentity}.  
Here we note that the covariance $\covCntrVec$ of the ensemble members controls the degree of satisfying the tangency approximation, and its update every iteration is influenced by the Hessian, or the local curvature of $\cost$. Large deviations of output $\obsMat$ (high sensitivity) implies large local curvature, which will subsequently shrink control-vector variances. Although this effect will improve gradient accuracy, since subsequent ensemble members are sampled closer together, covariance shrinkage can degrade convergence rates. Shrinking control parameter variances implies increased trust in them. Thus, the second term in (\ref{eq:costLinearization}) outweighs the first term and data is ignored. The optimizer may appear to identify a local optimum. 

In order to mitigate over shrinking control-vector covariances, \cite{anderson1999monte} recommended multiplicative inflation, which we implement by ${\pertMat}_{i+1}=\sqrt{A}\pertMat_i$ and thus ${\covCntrVec}^{\!\!\!\scriptscriptstyle{{i+1}}}= A/(N-1){\pertMat}_{\scriptscriptstyle{{i}}}{\pertMat}_{\scriptscriptstyle{{i}}}^\transpose$. Covariance inflation guards against prior distributions becoming so narrow that observation data are essentially ignored, halting reduction in $\cost$.  When the cost reduction is relatively weak, we can apply covariance inflation during the EnVar procedure. Appendix \ref{app:lorenz} analyzes the effect and shows improvement when used in conjunction with optimal sensor weighting. The EnVar approach is described in algorithm \ref{alg:EnVar}. 

\begin{minipage}{1\textwidth}
\begin{algorithm}[H]
\SetAlgoLined
Iteration $i=0$\\
 Estimate the initial control vector $\mathbf{c}_{\scriptscriptstyle{{0}}}$\\
\mbox{Generate ensemble members by sampling the normal distribution~$N(\mathbf{c}_{\scriptscriptstyle{{0}}},\covCntrVec)$}\\
 Acquire mean model observations $\mathcal{M}(\mathbf{q}_o)$ \\
Evaluate the initial cost $\cost_o$\\
 \While{Insufficient reduction in cost}{
  \mbox{Acquire ensemble member model observations: $\mathcal{M}(\mathbf{q}^{\scriptscriptstyle{(j)}})~~|~~j=1\ldots \Nens$} \\
  Construct observation matrix $\mathbf{H}$\\
  Compute the gradient $\nabla\LinCost$  and Hessian $\hessian$\\
  Acquire optimal weights \\
  Construct the next ensemble mean $\mathbf{c}_{i+1}$\\
  Acquire mean model observations $\mathcal{M}(\mathbf{q}_{i+1})$ \\
  Compute the cost $\cost$ and check convergence criteria\\
  Generate ensemble members about $\mathbf{c}_{i+1}$ with posterior statistics\\
  Update the covariance matrix $\covCntrVec$\\
  \If{Condition for covariance inflation met}
  {
    Amplify ensemble member perturbations by $\sqrt{A}$\\
    Update the covariance matrix $\covCntrVec$
  }
  $i=i+1$
 }
 \KwResult{Optimal inflow amplitudes and phases $\mathbf{c}_{\mathrm{opt}}=\mathbf{c}_{i}$}
\caption{Observation-infused EnVar technique}
 \label{alg:EnVar}
\end{algorithm}
\end{minipage}

\section{Simulation and optimization configuration}\label{sec:simConfig}
The ultimate future objective is to apply the observation-infused framework to experimental measurements. However, in order to fully assess performance of the method, we test our framework on observations generated from a precursor simulation. In particular, this substitute facilitates precise control of the observation error, access to the true inflow disturbances, and assessments beyond on the measurement domain. When needed, the implications of using simulation-generated observations in place of experimental measurements, e.g.\,perfect versus contaminated observations, will be indicated. In our presentation of the observation-infused approach, we maintain the perspective of having access to measurements only, as in an authentic prediction, and make comparisons with the `unknown' inflow and true flow field in section~\ref{sec:error} after the EnVar optimization is completed.

For all simulations, including the precursor DNS, the computational model solves the compressible flow equations in Cartesian coordinates (\ref{eq:mass}-\ref{eq:energy}) for the zero-pressure gradient flat-plate boundary layer, shown in figure~\ref{fig:configuration}~(a). 
The gas is assumed ideal with ratio of specific heats $\gamma=1.4$ and temperature-dependent viscosity following Sutherland's formula. 
The flow equations are solved using a standard fourth-order Runge--Kutta scheme and interior sixth-order finite differences. Near boundaries, stencils are biased and accuracy is reduced. 
In the presence of shocks, a fifth-order weighted essentially non-oscillatory scheme (WENO) is used locally. Detailed characterizations of the flow solver are provided elsewhere~\citep{larsson2009direct}.
The free-stream Mach number is $M_\infty = 4.5$, and the inflow Reynolds number is $Re \equiv \rho_\infty L U_\infty / \mu_\infty = 1800$ based on the length scale $L=\sqrt{\nu_\infty x / U_\infty}$. 
The computational domain size and grid resolutions are provided in Table \ref{tab:configuration}. 
In viscous units, the near-wall grid spacing is $(\Delta x^+, \Delta y^+, \Delta z^+)=(4.4, 0.25, 4.2)$.

\begin{table}\label{ref:tableError}
\begin{center}
\begin{tabular}{l c c c c c c c c  c c c}
Case & $M$ & $\sqrt{Re_{x_0}}$ & $L_x$ & $L_y$ & $L_z$ & $N_x$ & $N_y$ & $N_z$  & $ \measHorizon$ & $\measFrequency$ & $\Delta t$ \\\hline
DNS & 4.5 & 1800 & 2553 & 200 & 50 & 1703 & 200 & 36 & 349 & 0.873 & 0.174\\
LPSE & 4.5 & 1800 & 2100 & 200 & 50 & 85& 200 & 36  & 349 & 0.873 &--\\
\hline 
\end{tabular} 
\vspace*{-6pt}
\caption{Simulation and data acquisition parameters.}
\label{tab:configuration}
\end{center}
\end{table} 

\subsection{Precursor simulation and observations}\label{sec:dataacquisition}

In practice, the geometry of the test article and the spatial distribution and type of sensors (e.g.,~thermocouples, PCB sensors, etc.) are known. Also another sensible expectation, the nominal flow conditions are assumed to be known, including the density $\rho_\infty$, temperature $T_\infty$, viscosity $\mu_\infty$, and relative velocity $U_\infty$. Depending on the wind tunnel, a statistical characterization of the free-stream disturbances may have been made beforehand.  For flight tests, a sparse historical record of atmospheric disturbances is available, but the precise state depends on the time of day and year, global position, and altitude. \citet{schneider2001effects} reviews some of these relevant free-stream measurements for high-speed flight and ground-based tests; however, in the present study, no prior environmental disturbance data or physical constraints are used in the observation-infused framework.

{In order to generate the surrogate observations, the precursor simulation used the nonlinearly most dangerous inflow disturbance with a prescribed energy level for Mach 4.5 boundary layer~\citep{jahanbakhshi_zaki_2019}.
This configuration should be contrasted to the classical scenarios where either subharmonic or fundamental oblique waves are introduced in order to initiate secondary instability of a primary two-dimensional second mode. Transition due to the nonlinearly most dangerous disturbance is more challenging to predict for the following reasons: 
{(i) The inflow is comprised two planar second modes and an oblique first mode, which do not satisfy the fundamental or subharmonic resonance criteria}.  (ii) The energy in these modes at the inflow is sufficiently high that nonlinear interactions are relevant early upstream;  (iii) A series of nonlinear energy transfers spur new instability waves, not present at the inflow, and hence predicting the inflow by decoding downstream observations is more difficult than in classical fundamental and subharmonic transition; (iv)  Transition takes place most swiftly, while all other inflow disturbances would relatively delay transition. }
The nonlinearly most dangerous inflow condition is the `unknown' truth that we seek to predict.  It is noteworthy that, even in the context of the precursor simulation, the sensitivity of discrete wall pressure measurements and transition to the inflow are not known.   The EnVar algorithm, by attempting to discover the inflow condition, decodes the inflow mode amplitudes and phases that are pertinent for reproducing observations.  

For the surrogate measurements $\meas$, time traces of surface pressure are acquired from a regular array of 64 PCB-like sensors, $N_x^s\times N_z^s=8\times8$, which is a practical number used in ground-based~\citep{marineau2015investigation,zhu2016transition} and flight tests~\citep{thiele2018instrumentation}.
The sensor spacing is uniform in the span due to homogeneity of the flow in that dimension. 
In the streamwise direction, physical limitations often preclude placement of the sensors far upstream. In addition, upstream sensors may suffer low signal-to-noise ratio. {The $N$-factors for the second mode instabilities can provide some guidance, for example to cluster the sensors at the positions where the $N$-factors peak, although the uncertainty of transition location motivates a uniform spacing rather than clustering. Strategies for optimal sensor placement (see e.g.,~\cite{mons2019kriging}) in uncertain environments are the subject of ongoing research.  For the present purposes, and taking into account the above considerations, we distributed the sensors uniformly between $2000 \leq \sqrt{Re_x} \leq 2650$.}
  

\begin{figure}
    \centering
    \includegraphics[width=\textwidth]{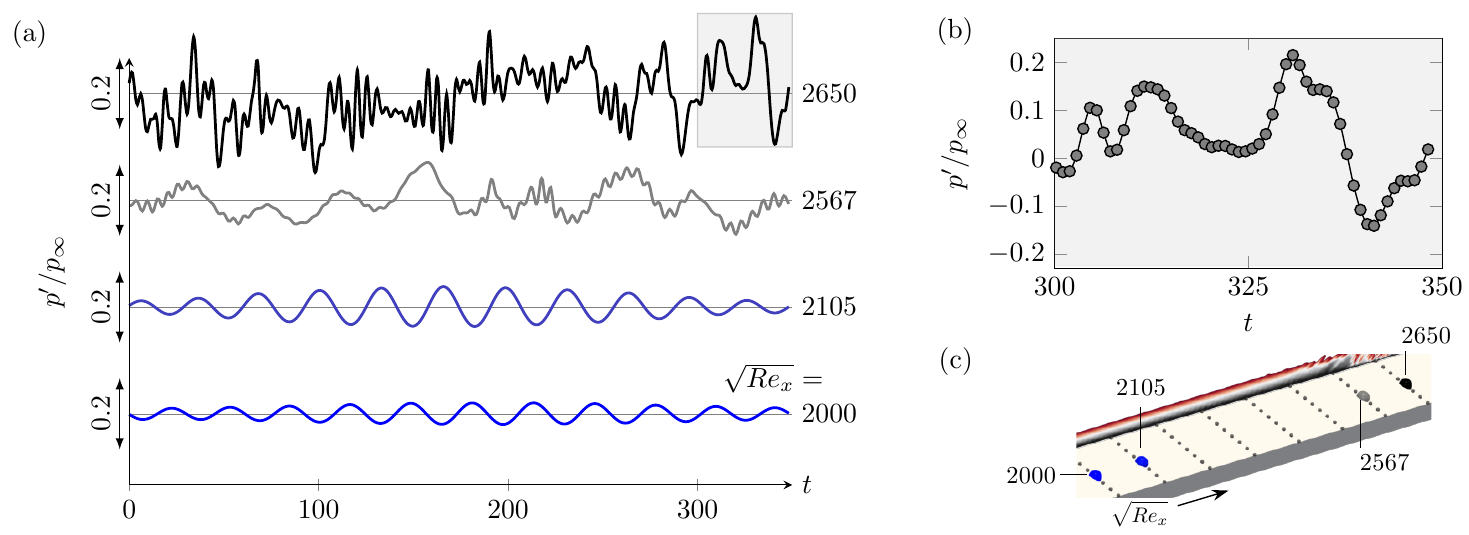}
    \caption{(a) Sample wall pressure signals from the precursor DNS, extracted at $z=20.8$ and $\sqrt{Re_x}=\{2000, 2105, 2567, 2650\}$. Horizontal lines indicate $p'=0$. (b)~Detail of the shaded region in (a); the acquisition frequency is shown by symbols. (c)~Detail of sensor positions: large symbols correspond to traces shown in a) and small symbols are remaining sensors.}
    \label{fig:pressureTraces}
\end{figure}

Once the flow has reached a statistically stationary state, observations were acquired for a duration $\measHorizon=349 L/U_\infty$, which was resolved using $N_t=400$ samples. While in physical experiments the observations may span longer duration, even when bound by the tunnel run time, the assimilation time horizon is shorter for computational efficiency since each EnVar iteration required $\Nens+1=17$ forward simulations. 
The horizon is however sufficiently long to capture an adequate number of waves, and in our case, it spans $\gtrsim 10$ periods of the second mode.  
In terms of sampling rate, in physical experiments it is a limitation of the sensor.  This reality was emulated by sub-sampling the simulation but ensuring that we resolve potential second-mode instabilities and nonlinearly generated harmonics.
Figure~\ref{fig:pressureTraces}~(a) shows traces from probes along the fourth streamwise ray of sensors $z=20.8$, and (b) shows the resolution of the acquisition sampling. 
The time dependent wall pressure from the sixty-four discrete sensors are then concatenated to form the observation vector, 
\begin{equation*}
    \meas=\left[p\left(\mathbf{x}^s_{1,1},t_1\right),\ldots, p\left(\mathbf{x}^s_{1,1},t_{N_t}\right),\ldots, p\left(\mathbf{x}^s_{\scriptscriptstyle{N_x^s,N_z^s}},t_{1}\right),\ldots,p\left(\mathbf{x}^s_{\scriptscriptstyle{N_x^s,N_z^s}},t_{N_t}\right)\right]^\transpose\!\!,
\end{equation*}
where $\mathbf{x}^s_{i,j}$ is the $(x,z)$ position of the $(i,j)$ sensor on the wall.

In each step of the EnVar algorithm, model observations $\projCntrVec$ are collected and compared to $\meas$. 
This step is performed similarly to the above description, where the wall pressures are recorded at the locations of the sensors and at the same sampling rate.  
Note that in experiments, physical sensors may introduce additional effects, for example filtering due to finite sensor size \citep{kennedy2018investigation}. For PCB probes, corrections to the spectra have been developed \citep{corcos1963resolution,lueptow1995transducer} that can be applied to $\meas$ directly.  Alternatively, the impact of the sensor can be incorporated in $\projCntrVec$.

\subsection{EnVar configuration}
Generally, environmental disturbances are a mixture of acoustic, vortical, and entropic components~\citep{kovasznay1953turbulence}, and in high-speed tunnels the makeup of disturbances is dominated by intense acoustic waves \citep[e.g.][]{pate1980effects}.  In the present simulation domain, we express the unknown inflow disturbance as a superposition of instability waves of the boundary layer.  This choice is natural given the exhaustive literature on boundary-layer receptivity and transition mechanisms that are framed in terms of originally linear instability waves~\citep{mack1975linear,malik1990numerical,fedorov2011transition}.
Despite our particular parameterization, the EnVar technique is amenable to other forcing schemes and can also accommodate additional uncertain parameters (e.g. surface roughness, real-gas effects, etc.).

The full inflow condition is expressed as a superposition of the mean state and instability waves, 
\begin{align}\label{eq:inflowSuperposition}
    \mathbf{q}_o&=[\overline{\rho}, \overline{u}, \overline{v}, \overline{w}, \overline{T}](y) \\\nonumber&+ Re\left(\sum_m\sum_f c_{\scriptscriptstyle{{f,m}}} [\hat{\rho},\hat{u},\hat{v},\hat{w},\hat{T}]_{\scriptscriptstyle{{f,m}}}(y) \exp[\mathrm{i}\beta_m z-\mathrm{i}\omega_f t + \mathrm{i} \theta_{\scriptscriptstyle{{f,m}}}] \right).
\end{align}
In this form, the control vector is comprised of the unknown amplitudes and phases $\mathbf{c}=[\ldots,c_{\scriptscriptstyle{{f,m}}},\theta_{\scriptscriptstyle{{f,m}}},\ldots]$ of the inlet instability waves.
{The choice of the range and the resolution of the inflow frequencies is guided by preliminary spectral analysis of the observations, which is the natural approach to adopt in practice with experimental measurements.} 
Initially, thirty-three wavenumber-frequency pairs are considered, which encompass first and second mode instabilities for the range $1800 \leq \sqrt{Re_x} \leq 2650$. 
The nondimensional frequencies and integer spanwise wavenumbers are
\begin{align}
F&=\frac{\omega_f}{\sqrt{Re_{x_o}}}\times 10^6 ~~\text{for}~f\in[1,2,\ldots, 11]~\text{and}\label{eq:Fomega}\\
k_z&=  \frac{\beta_m L_z}{2 \pi}= m~~\text{for}~m\in[0,1,2].
\end{align}
For each $(F,k_z)$ pair, we consider the most unstable slow modes~\citep{fedorov2011transition} at the inflow $\sqrt{Re_x}=1800$, from the discrete Orr–Sommerfeld and Squire spectra. Note that the choice of normalizing mode shapes (e.g.\,$\hat{\rho}$) is not unique and can influence reproducibility.
We adopt the energy normalization \citep{mack1984boundary,hanifi1996transient}, 
\begin{equation}
    \mathcal{E}_{\scriptscriptstyle{{f,m}}} = \frac{1}{2} \int_0^{L_y} \hat{\mathbf{q}}^\star \mathbf{M} \hat{\mathbf{q}}~dy, 
\end{equation}
where ($^\star$) is the complex conjugate transpose, $\hat{\mathbf{q}}$ is the solution vector $\left[\hat{\rho},\hat{u},\hat{v},\hat{w},\hat{T}\right]_{\scriptscriptstyle{{f,m}}}^\transpose$, and the matrix $\mathbf{M}$ scales the density, momentum, and internal energy by
\begin{equation}\label{eq:weightSolution}
    \mathbf{M} = \mathrm{diag}\left[\frac{1}{\gamma(\gamma-1)M^2} \frac{\overline{T}}{\overline{\rho}},\quad \overline{\rho},\quad\overline{\rho},\quad\overline{\rho},\quad \frac{1}{\gamma(\gamma-1)M^2} \frac{\overline{\rho}}{\overline{T}} \right],
\end{equation}
which includes the non-dimensionalization of our equations.

With all thirty-three instability waves considered, the control vector would be comprised of sixty-six unknowns.  In order to reduce the dimensionality of the EnVar optimization, it is desirable to reduce the length of the control vector and to establish a good initial guess.  Both objectives are achieved using a LPSE estimate of the control vector, which is discussed in \S\ref{sec:pse}. The eight instability waves that are predicted by the LPSE estimation to have the largest inflow amplitudes will form the control vector for subsequent iterations using the nonlinear EnVar algorithm.  Therefore, the number of ensemble members in the main algorithm was $\Nens=16$, which is the same size as the control vector.  Each EnVar iteration thus involves $\Nens+1=17$~simulations, where the additional DNS corresponds to the mean of the ensemble.  

The initial covariance matrices are $\covMeasVec=\sigma_m^2 \identity$ where $\sigma_m=10^{-1.75}$ based on the intensity of the recorded pressure fluctuations ($p'_{rms} \approx 0.017$  at the first sensor row $\sqrt{Re_x}=2000$), and $\covCntrVec=\sigma_c^2\, \text{diag}[\mathbf{c_o}]^2$ where $\sigma_c=0.4$. A preliminary test was conducted which showed that the relatively large value of $\sigma_c$ yielded variation of outcomes across all of the sensor positions which was deemed sufficient for starting the EnVar procedure. Had the ensemble members yielded indistinguishable observations, perhaps from a very tight $\covCntrVec$, the observation matrix would be nearly singular affecting optimization. Although a relatively large $\sigma_c$ may violate the tangent approximation, initially, the $\covCntrVec$ is iteratively refined by exploiting the local shape of the cost function using (\ref{eq:pertMatPosterior}--\ref{eq:covariancePosterior}). 

\subsection{Initial estimate of the control vector}
\label{sec:pse}

The EnVar algorithm requires an initial estimate of the control vector, $\mathbf{c}_{\scriptscriptstyle{0}}$.
One may assume a uniform distribution of energy among all possible inflow instability waves, or a turbulent energy spectrum, or any other condition motivated by prior knowledge of the flow environment.  
Since the first estimate $\mathbf{c}_{\scriptscriptstyle{0}}$ has bearing on performance of the gradient-based optimization, it warrants careful consideration.
Here, we present a method reliant on the linear parabolized stability equations (LPSE).  The motivation for this choice is that it provides a viable physical explanation of the observations, within well established set of assumptions.  This interpretation is valuable in itself when computational resources are not available to solve the full nonlinear problem.  In addition, when sensors are in the linear disturbance regime\textemdash an assumption that we do not adopt in this work\textemdash the accuracy of this inflow estimate is fully determined by the accuracy of the PSE approximation and the measurement uncertainty.

Due to linearity of PSE, each inflow instability wave evolves independently and can be considered as an ensemble member.
Instead of the Navier--Stokes model, we express the LPSE as $\mathbf{q}=\mathcal{L}(\mathbf{c})$. The cost functional is then, 
\begin{equation}\label{eq:linearCostPSE}
    \cost = \frac{1}{2}\norm{\meas - \mathbf{L}\mathbf{v} }^2 + \frac{\varrho}{2} \norm{\mathbf{v}-\mathbf{v}_o}^2.
\end{equation}
The columns of the matrix $\mathbf{L}$ have the form, 
\begin{equation}\nonumber
    \mathbf{L}=\left[~\ldots~|~\ldots~|~~\mathrm{Re}\left(\hat{p}_{\scriptscriptstyle{{f,m}}}\right)~~|~\, -\mathrm{Im}\left(\hat{p}_{\scriptscriptstyle{{f,m}}}\right)~~|~\ldots~|~\ldots~\right], 
\end{equation}
with each pair being the real and imaginary parts of the Fourier transform of the observed pressure due to mode $(f,m)$.  
The new control vector is therefore $\mathbf{v}$ whose  elements $\mathbf{v}=\left[\ldots, v_{\scriptscriptstyle{{f,m}}},w_{\scriptscriptstyle{{f,m}}},\ldots\right]^\transpose$ are the unknown modal coefficients. 
{The second term in (\ref{eq:linearCostPSE}), weighted by a scalar $\varrho$, accounts for prior information $\mathbf{v}_o$  regarding the disturbance environment, for example from characterization of the free stream conditions in a tunnel or in flight. Even absent such knowledge $(\mathbf{v}_o=0)$, the second term serves to regularize the control vector such that we identify the inflow condition $\mathbf{v}$ that best reproduces the observations while also having a small quadratic norm.}
Since (\ref{eq:linearCostPSE}) is quadratic, the minimizer of $\cost$ is obtained by taking the gradient and assuming stationarity, which yields, 
\begin{equation}\label{eq:leastsquaresPSE}
    \mathbf{v}=\left(\mathbf{L}^\transpose \mathbf{L} +\varrho \mathbf{I}\right)^{-1}\left(\mathbf{L}^\transpose\mathbf{m}^\dagger + \varrho \mathbf{v}_o \right).
\end{equation}

The estimate of the control vector (\ref{eq:leastsquaresPSE}) can use a subset of sensor data ($\mathbf{m}^\dagger$) that are best suited for the linear approximation of PSE. This step benefits from examination of the spectrum of the observations (see figure~\ref{fig:spectra}), which show the presence of fundamental frequencies, but also the signature of higher harmonics before becoming more broadband by the fourth sensor row.  For our configuration, we used observations from the first row of sensors only, $\meas^\dagger=\meas~|~\sqrt{Re_x}=2000$, and the parameters of the PSE computations are summarized in table \ref{tab:configuration}.  {For our sensor network, the LPSE estimation yields a good estimate of the inflow condition without regularization ($\varrho=0$), although we caution that this is not generally the case especially for a small number of sensors. For example, given only one sensor, waves with the correct frequency but any spanwise wavenumber $k_z$ can be phase-shifted and scaled to precisely match the data, yielding an infinite number of potential solutions and (\ref{eq:leastsquaresPSE}) becomes ill-conditioned. In that case, include $\varrho$ ensures that the LPSE estimation selects the mode with the lowest quadratic norm.}

Once $\mathbf{v}$ was acquired, we identified the eight largest modal amplitudes in the ranges $40\leq F\leq 110$ and $0\leq k_z\leq 2$. 
Only the associated instability waves were retained in the subsequent EnVar optimization where the initial estimate of the control vector was
\begin{equation*}
    \mathbf{c}_o=\left[\ldots,{\sqrt{v^2_{\scriptscriptstyle{{f,m}}}+w^2_{\scriptscriptstyle{{f,m}}}}},\, {\tan^{-1}(w_{\scriptscriptstyle{{f,m}}}/v_{\scriptscriptstyle{{f,m}}})},\ldots\right]^\transpose.
\end{equation*}
%

\section{Results}\label{sec:results}

\begin{figure}
        \centering
        \includegraphics[page=1,width=0.99\textwidth]{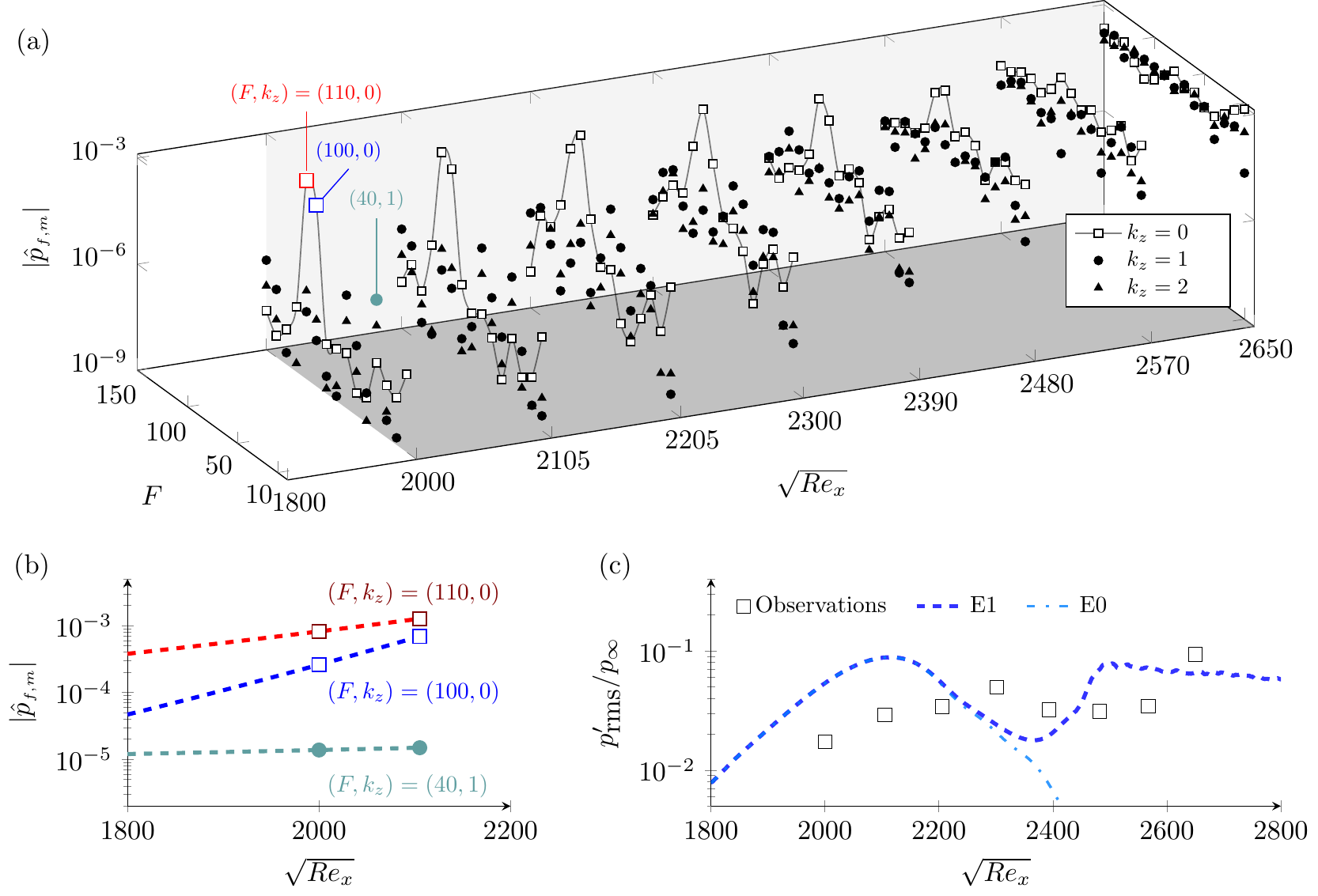}
        \caption{(a) Pressure spectra from the 64 sensors placed on the adiabatic wall $2000\leq \sqrt{Re_x}\leq 2650$.
        Three modes with largest amplitudes at $\sqrt{Re_x}=2000$ are identified.  
        (b) Possible inflow conditions using least-squares fit of amplitudes of those three modes at the first two sensor rows. 
        (c) Wall pressure prediction from DNS using the inflow conditions E0 and E1.}
        \label{fig:simpleInterpretation}
\end{figure}

Before reporting the predictions by the EnVar algorithm, we examine the observations, or pressure traces, without aid of the equations and attempt to identify information that may be pertinent to the transition mechanism.  In particular, figure~\ref{fig:pressureTraces}~(a) shows what appears to be a single dominant instability wave at $\sqrt{Re_x}=2000$ and broadband $p'$ signal at $\sqrt{Re_x}=2650$.  However, Fourier analysis uncovers a richer spectrum of waves, including the signature of  three-dimensional ones (figure~\ref{fig:simpleInterpretation}~a).  The peak two-dimensional instabilities, $F=100$ and $110$, amplify between $2000\leq\sqrt{Re_x}\leq2300$, and the signal appears to become broadband by $\sqrt{Re_x}\geq 2450$. Based on the figure and also linear PSE of the potential instability waves (performed to identify the initial guess of the EnVar algorithm), transition would be ascribed to amplifying second-mode instabilities; yet with limited information, the role that lower amplitude modes play in spurring key interactions for transition is uncertain, in fact unknown.

{We will consider two estimates of the inflow condition using only the data, akin to common interpretations of experimental measurements. The purpose is to provide the backdrop for evaluating the physics-based EnVar approach. The modes that have the largest wall pressure $|\hat{p}_{f,m}|$ at $\sqrt{Re_x}=2000$ are isolated and their streamwise development is plotted in figure~\ref{fig:simpleInterpretation}~(b). The first two sensor rows suggest exponential amplification model,}
\begin{equation}
    \ln|\hat{{p}}_{\scriptscriptstyle{{f,m}}}|(x)= a_o + a_1 x. 
\end{equation}
{This interpretation and similar approaches in the literature are predicated on linear assumptions, namely that instability waves evolve independently and amplify exponentially.}
{Using the above exponential model, we can infer the inlet amplitude from the sensor data.  The first estimate of the inflow (case `E0') is comprised of the two planar waves $\modeFkz{100}{0}$ and $(110,0)$, since measurements often focus on identifying planar second modes. The second estimate of the inflow (case `E1') also includes the oblique wave $\modeFkz{40}{1}$.  
In addition, for both E0 and E1, we assigned 1\% of the total inflow disturbance energy $\mathcal{E}_o$ to all the remaining instability waves, including three-dimensional ones.} Table~\ref{tab:energy} summarizes the distribution of energy for these cases.



\begin{table}
\centering
\begin{tabular}{ lccccc } 
 \multirow{2}{*}{Case} & \multirow{2}{*}{$\mathcal{E}_o\times10^5$} & \multicolumn{4}{c}{$\mathcal{E}_{(F,k_z)}/\mathcal{E}_o~(\%)$}\rule[-1.5ex]{0pt}{0pt} \\  \cline{3-6}
      &   & (100,0)\rule{0pt}{2.75ex} & (110,0) & (40,1) & other modes \\ \hline
  E0  & 5.7 &  ~2 & 97 & -- &  1  \\ 
  E1  & 6.8 &  ~2 & 80 & 17 &  1  \\ 
Truth & 2.0 &  31 & 26 & 43 & --  \\ \hline
\end{tabular}
\caption{Distribution of modal energy.}
\label{tab:energy}
\end{table}

Figure \ref{fig:simpleInterpretation}~(c) shows the outcome of simulations that adopt E0 and E1 as inflow conditions.  Both interpretations are incongruous with the root-mean-square (rms) of the observed wall pressure. In both cases, the rms pressure fluctuations are over-predicted by almost a factor of three, near $\sqrt{Re_x}\approx 2100$. {In case E0, transition is entirely absent within the computational domain even though we distributed 1\% of the energy in various modes to promote secondary instability of the planar waves.  When mode $\modeFkz{40}{1}$ is included (case E1), transition is premature. Worse, the cause of discrepancy is unknown and remains speculative.}


Unlike the above empirical approach, we wish to minimize ambiguities of transition mechanisms and contradicting interpretations. Our approach, the observation-infused EnVar framework, seeks the unknown inflow that optimally reproduces the observations.

\subsection{Optimal interpretation from the observation-infused framework}

\begin{figure}
    \centering
    \includegraphics[width=1.\textwidth]{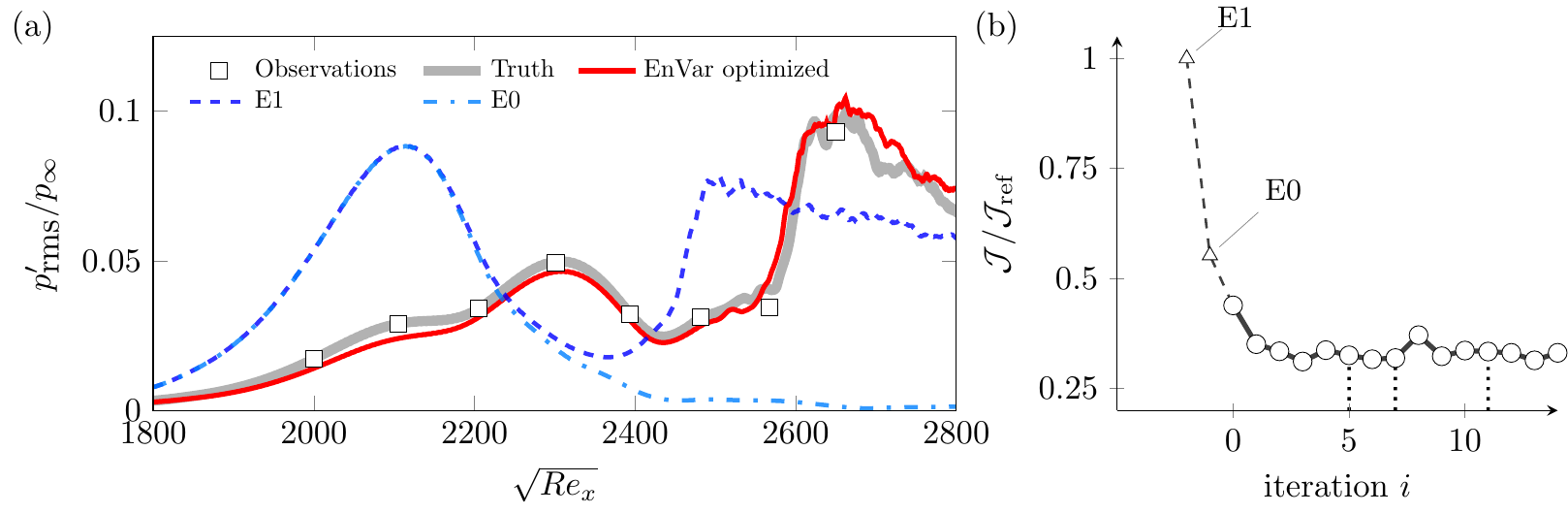}
    \caption{(a) Intensity of wall pressure fluctuations. 
    (b) The normalized cost function; vertical dashed lines identify iterations where $\Sigma_c$ was inflated.}
    \label{fig:rms-cost}
\end{figure}

The optimization seeks inflow amplitudes and phases which minimize the discrepancy between the simulated and observed wall pressures, acquired from a regular array of 64 PCB-like sensors. Figure~\ref{fig:rms-cost}~(a) contrasts the wall pressure root mean square (rms) between the predictions and observations. Unlike the predictions assuming exponential amplification of modal amplitudes {(cases E0 and E1)}, the EnVar technique identifies an inflow condition that accurately reproduces the sensor data. In fact, the observation-infused simulation tracks the pressure development even upstream, in between, and downstream of the sensors and predicts transition location, far outperforming empirical interpretations. 

Although, the rms is a useful statistical comparison, the optimization exploited point-wise, time-resolved observations in order to reconstruct, as close as possible, the exact state-space trajectory of the experiment. Point-wise deviations are quantified by the cost function $\cost$ using (\ref{eq:cost}), which is reported in figure~\ref{fig:rms-cost}~(b) for E0 and E1, and for all iterations during the EnVar gradient-based optimization. In concurrence with figure~\ref{fig:rms-cost}~(a), the EnVar approach improves the interpretations, although the reduction of the cost function by a factor of three appears humble relative to the improved prediction of the root-mean-square pressure fluctuations.  

The small reduction of the cost functional could be a symptom of reaching a local optimum. An obvious factor is the initial estimate of the control vector, which in our case is from the PSE solution and is hence physical.  Another key element is the landscape of the cost functional $\cost$ that is being explored by the EnVar technique.  The representation of the landscape is partly determined by the outcomes within the observation matrix $\obsMat$, because the variance of the ensemble members at each iteration depends on the local Hessian $\hessian$. When the surface of $\cost$ is highly oscillatory, the variance of the ensemble reduces, the members explore a smaller region, and reduction in $\cost$ ceases.  
In order to mitigate this potential issue, we have inflated the covariance of the control vector at iterations $i=5$, $i=7$, and $i=11$, which were selected based on the flattening of $\cost$.  
Despite the covariance inflation, the reduction rate of the cost function appears inappreciably affected.
Another important factor that impacts the optimization is the shape of the cost function, which depends on the placement of sensors, their relative weighting, the relative trust in observations versus in the prior control vector.  
While we will not vary the placement of the sensors, we will evaluate the degree to which the present uniform sensor weighting impacts the optimization in \S\ref{sec:optimalWeighting}, and also using a model system in Appendix~\ref{app:lorenz}.  In particular, section~\ref{sec:optimalWeighting} will introduce a weighting strategy which will achieve over two orders of magnitude cost reduction in as little as one-third of the iterations. Tests indicate that discounting the observations from sensors that contribute to the largest uncertainty in the inverse problem, through chaos, reduces the oscillations in the landscape of the cost function and improves the EnVar solution.  Despite these allusions to enhancing success of EnVar, we proceed in demonstrating the best that we can achieve assuming all observations are weighted equally.

\begin{figure}
    \centering
    \includegraphics[width=1.\textwidth]{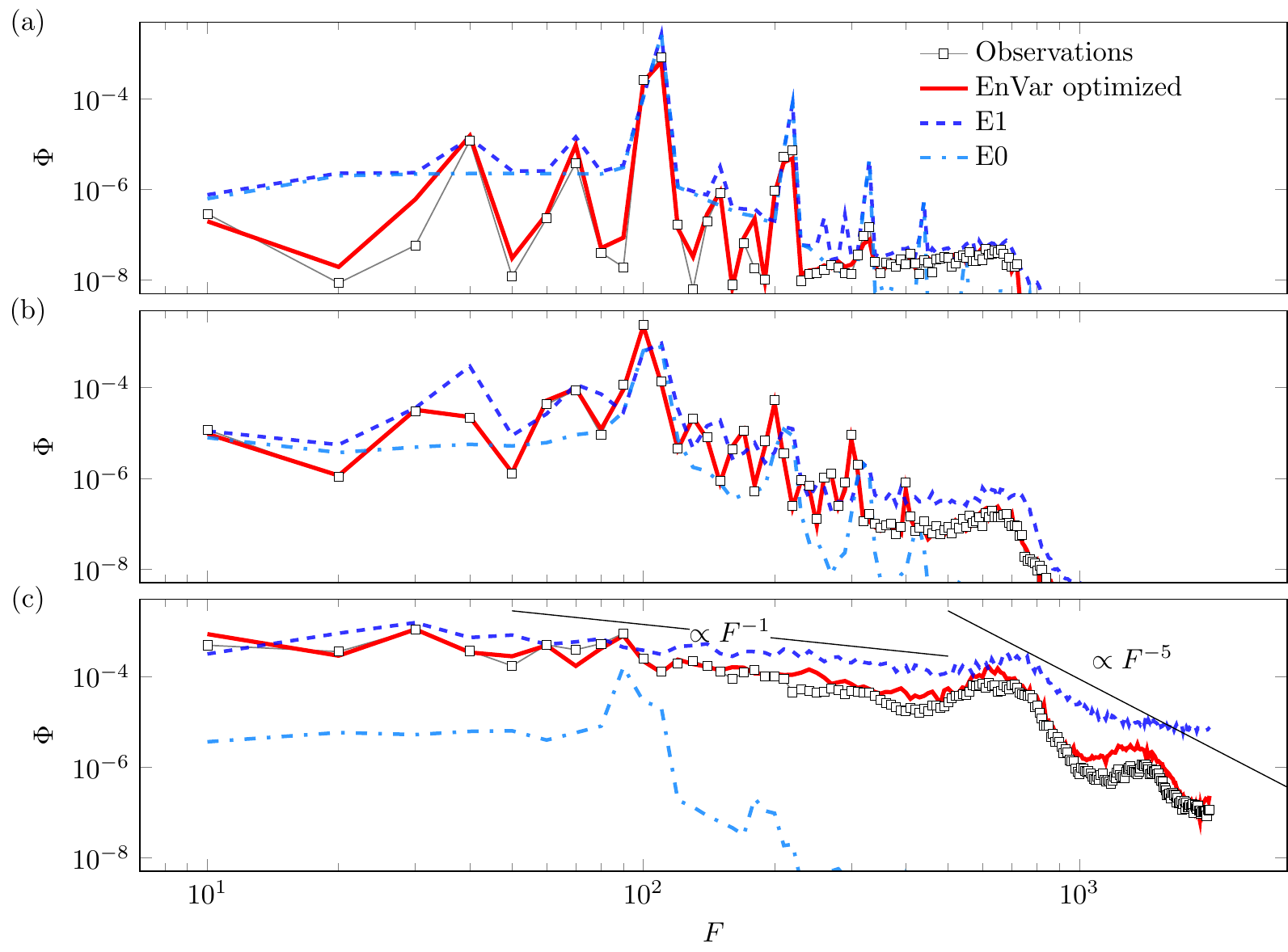}
    \caption{Wall pressure spectra averaged in $z$: (a)~first row $\sqrt{Re_x}=2000$ , (b)~fourth row $\sqrt{Re_x}=2300$, and (c)~seventh row $\sqrt{Re_x}=2570$. For reference, the $F^{-1}$ and $F^{-5}$ sloped lines indicate spectral roll off in equilibrium boundary-layer turbulence~\citep{bull1996wall}.}
    \label{fig:spectra}
\end{figure}

The pressure spectra provide a further refined view of the wall-pressure reconstruction, beyond the rms in figure \ref{fig:rms-cost}.  The spectra can, for example, indicate the presence of key instabilities, higher harmonics, and spectral broadening possibly due to nonlinearity. 
Figure~\ref{fig:spectra} shows the pressure frequency spectra averaged in the span across the $N_{z}^s=8$ sensors, 
\begin{equation}\label{eq:fft}
    \Phi = \frac{1}{N^s_{z}}\sum_{j=1}^{N_{z}^s} \bigg{|} \frac{1}{N_t} \sum_{n=0}^{N_t-1} p(\mathbf{x}^s_{i,j},t_n) \exp\left[-\mathrm{i} \left(2 \pi f / T_a\right) n \Delta t_a  \right]\bigg{|}
\end{equation}
where $f=[0,1,\ldots,N_t-1]$ is the integer frequency. Results are reported for the first, fourth, and seventh sensor rows, which respectively correspond to the early pre-transitional regime, the region near the peak rms wall pressure, and the region immediately preceding transition. The EnVar optimized solution reproduces the spectral content of the observations. In figure~\ref{fig:spectra}~(a) there is mild deviation for energy in at frequencies $F=[20, 30, 50, 90]$, \textit{albeit} relatively low amplitude, which suggests that the EnVar-estimated inflow spurs nonlinear interactions earlier than in the true flow. Yet by $\sqrt{Re_x}=2300$, the EnVar predictions are indistinguishable from the observations. In contrast, case E0 that overlooks mode $\modeFkz{40}{1}$ does not yield amplifying instabilities $F=[30,60,70]$, possibly due to nonlinear interactions, and transition to turbulence is not achieved within the simulation domain. Including $\modeFkz{40}{1}$ in the physics-free exponential model (case E1) can lead to transition, but the levels in the spectra disagree with the observations (figure~\ref{fig:spectra}~(c)), which demonstrates that the transition mechanism in such approach is inconsistent with the true flow.

\subsection{{Assessment of performance using the true state}}\label{sec:error}
\begin{figure}
     \centering
     \includegraphics[width=1\textwidth]{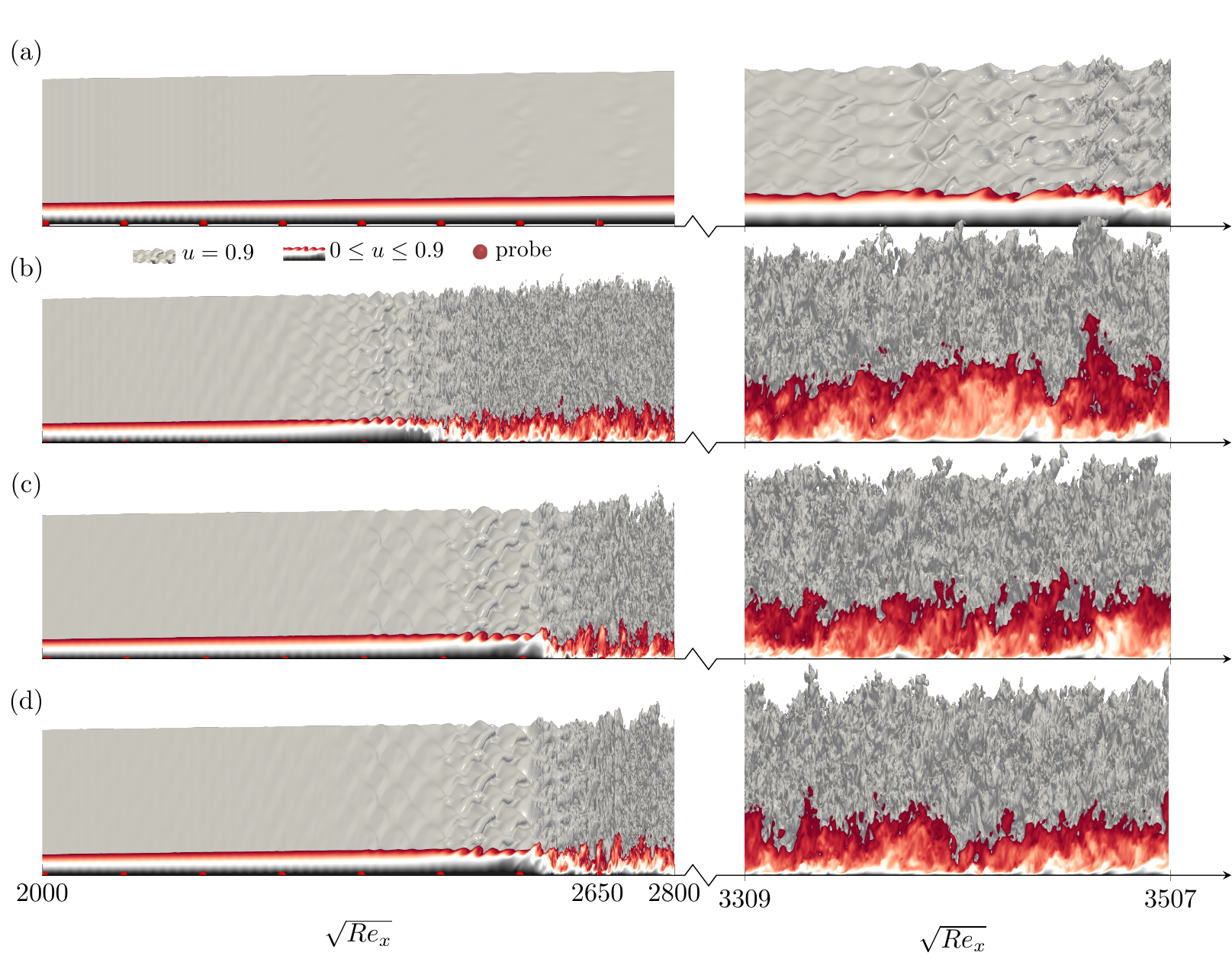}
     \caption{Visualization of streamwise velocity from DNS of (a) E0, (b) E1, (c) EnVar optimized inflow after 14 iterations, and (d) true inflow condition.}
     \label{fig:visualComparision}
 \end{figure}

To this point, the accuracy of the EnVar predictions was assessed against the discrete observations, which are a subspace of the complete flow state. 
However, since the observations were generated from a simulation as a surrogate for experiment, we also have access to the entire true state for the purposes of benchmarking the accuracy of our predictions.  In particular, transition in the true flow was caused by the nonlinear most unstable inflow comprised of modes $\modeFkz{40}{1}$, $\mode{100}{0}$, and $\mode{110}{0}$ as noted in \S\ref{sec:dataacquisition}.  We examine the details of the transition process beginning with flow-field visualizations in figure \ref{fig:visualComparision}. 
The physics-free interpretations, E0 and E1, which adopt an exponential fit of the sensor data, are shown in the first two panels; they are qualitatively different from the true state (panel $d$).  These results also exemplify the sensitivity of boundary layer transition to the influence of $(F,k_z)=(40,1)$: Lacking this particular mode (figure \ref{fig:visualComparision}$a$) yields persistently laminar field within the simulation domain;  Over-predicting the amplitude of that mode (figure \ref{fig:visualComparision}$b$) accelerates transition significantly. 
Relative to the physics-free interpretations in the first two panels, the EnVar optimized simulation in panel $c$ develops similarly to the truth. The formation of three-dimensional structures above the wall and transition are observed near $\sqrt{Re_x}\approx 2400$. Note that, as indicated in table~\ref{tab:configuration}, the flow region shown extends much farther downstream than the EnVar observations and also the domain adopted for EnVar optimization in order to assess the accuracy at long distances. In the turbulent regime, $3309\leq \sqrt{Re_x}\leq 3507$,  the instantaneous flow deviates from the truth due to the chaotic nature of turbulence which amplifies differences at the inflow.  

\begin{figure}
    \centering
    \includegraphics[width=1.\textwidth]{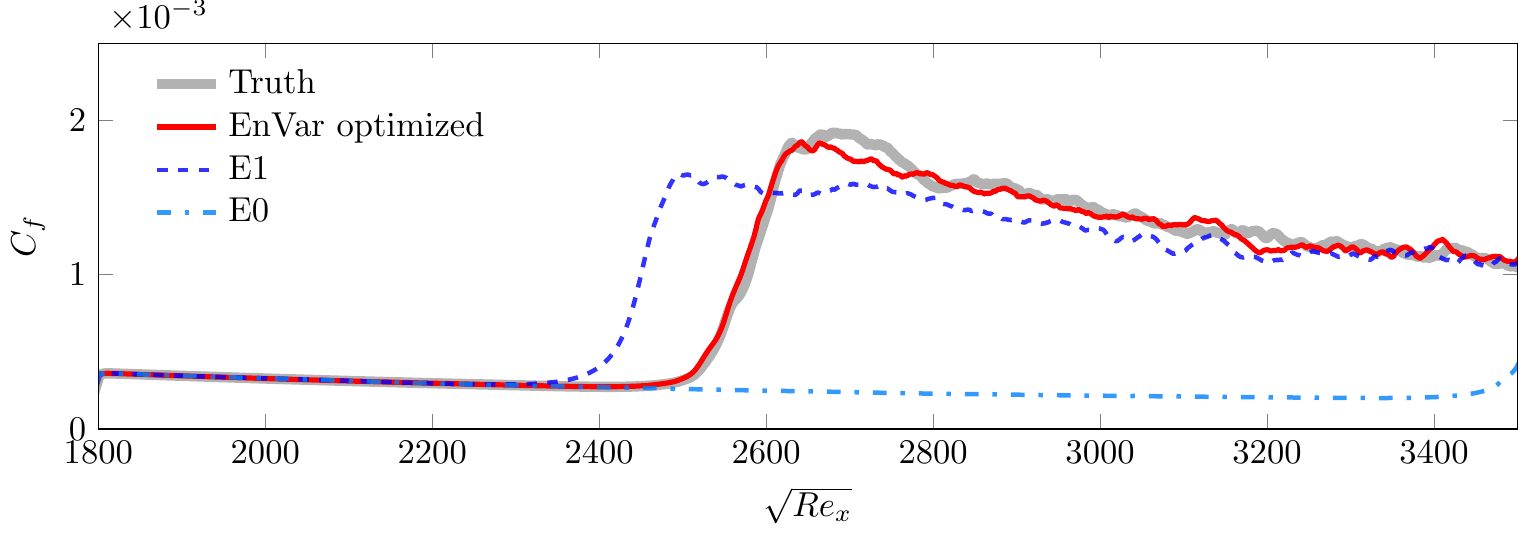}
    \caption{Downstream dependence of the mean skin-friction coefficient.}
    \label{fig:cf}
\end{figure}

An important measure of accuracy in predicting transitional flows is the ability to identify transition location.  
Figure~\ref{fig:cf} shows that the skin-friction coefficient, $C_f=\frac{2}{\rho_\infty U_\infty^2}\langle \tau_w\rangle_{zt}$ where $\langle \cdot \rangle_{\chi}$ denotes averaging in the ``$\chi$'' direction, has a similar level of agreement as the rms wall pressure in figure~\ref{fig:rms-cost}~(a)
Whether adopting wall shear-stress observations, instead of wall pressure, would yield less or more accurate predictions than in figure \ref{fig:cf} is not a simple question for two reasons: Firstly, as a practical consideration, the errors in acquiring observations depend on the sensing modality. We are reminded, however, that this issue is mitigated here because we adopted a precursor simulation as a proxy for experiments, and our observations are sampled in the exact same manner as in the EnVar procedure. Secondly, and perhaps more importantly, the landscape of the cost functional changes depending on the choice of observations. The accuracy of interpreting different measurement modalities and the impact of their respective uncertainties is the subject of ongoing work.

\begin{figure}
    \centering
    \includegraphics[width=0.99\textwidth]{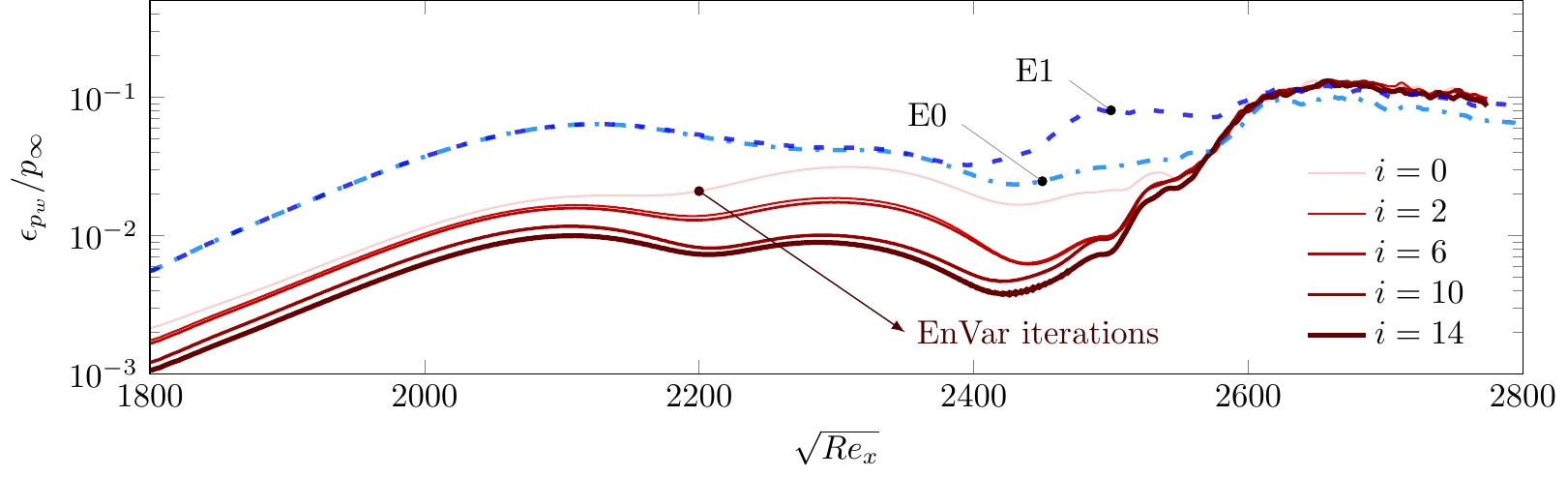}
    \caption{Wall pressure errors for inflow conditions estimated from (solid) successive EnVar iterations $i=\{0,2,6,10,14\}$, as well as exponential modal growth models (dash-dotted)~E0 and (dashed)~E1.}
    \label{fig:errorWallPressure}
\end{figure}

In addition to graphs demonstrating qualitative agreement between the EnVar-estimated and true flows, we quantify the deviation between flow fields.
We begin with the root-mean-squared error in the wall pressure, $\epsilon_{p_w}=\langle (p'_w - {p'_w}^{\!\!{\scriptscriptstyle{\top}}})^2 \rangle_{zt}$.
Figure~\ref{fig:errorWallPressure} shows that inflow conditions E0 and E1 that are exponential fits from downstream sensor data lead to large errors. 
In contrast, the EnVar technique provides the smallest wall-pressure errors, which concurs with the results in figure~\ref{fig:rms-cost}.
In the region $2100\leq Re_x\leq 2500$, nonlinear interactions lead to important energy exchanges among modes for this configuration \citep{jahanbakhshi_zaki_2019}. 
Within this region and also upstream, successive EnVar iterations progressively refine wall pressure predictions and reduce the errors, until the point of transition to turbulence. 
In the turbulence ($\sqrt{Re_x}\gtrsim 2600$), the errors are commensurate with the root-mean-square pressure signal, $\epsilon_{p_w}\approx {p’^{\scriptscriptstyle{\top}}_{\mathrm{rms}}}$.  Although the upstream flow was captured well by the EnVar optimization, to no surprise the streamwise variation of errors increase near the turbulence since small inflow deviation ($\approx 0.1\%$ wall pressure error at $\sqrt{Re_x}=1800$) can have a drastic impact on instantaneous trajectory of the flow as confirmed by flow visualizations in figures~\ref{fig:visualComparision}~(c,d). 

\begin{figure}
    \centering
    \includegraphics[width=0.99\textwidth]{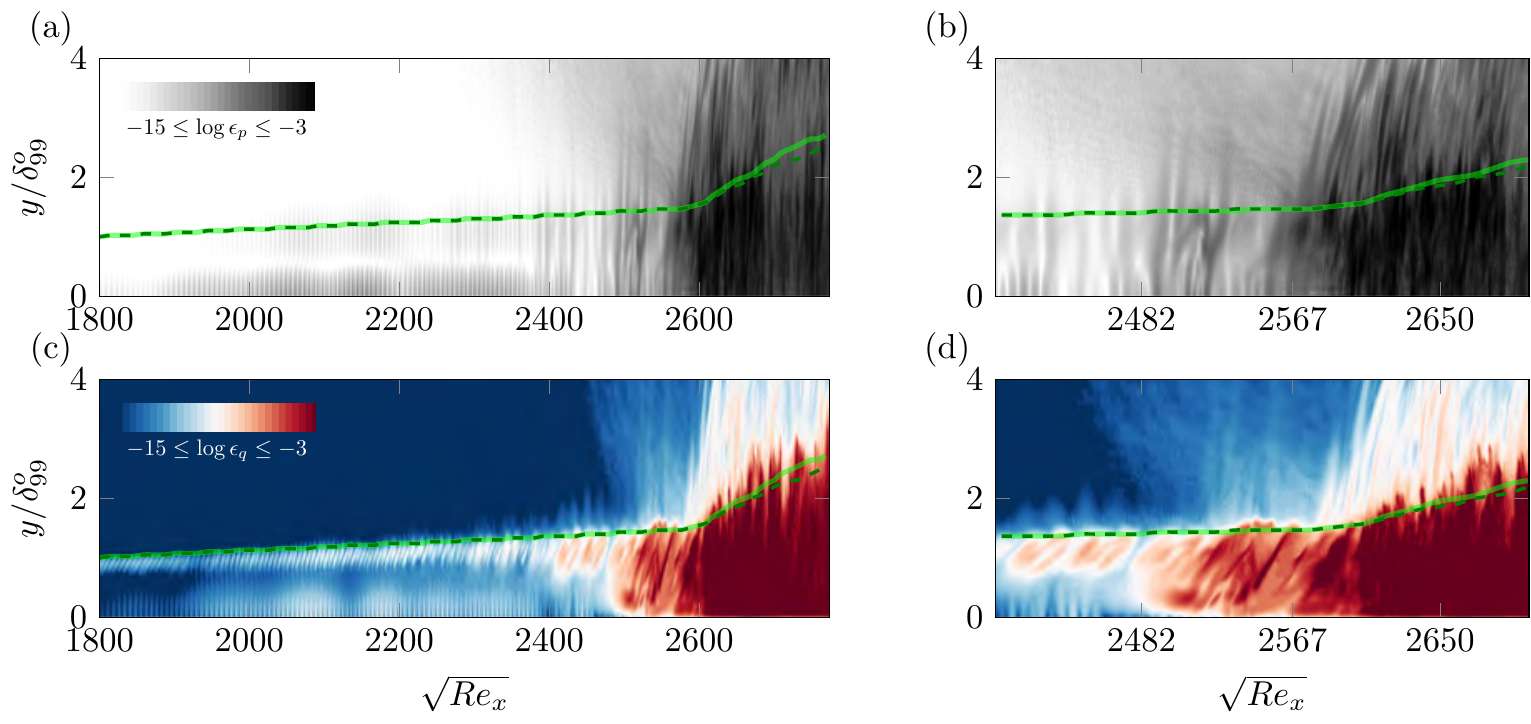}
    \caption{Instantaneous errors in EnVar estimated flow relative to the truth. Contours show spanwise average errors in (a,b) pressure and (c,d) full state vector. 
    Overlaid curves represent the position of $\delta_{99}(x)$ for the (solid) truth and (dashed) EnVar prediction.}
    \label{fig:errorfields}
\end{figure}

The above comparison was focused on errors along the wall, the plane where the sensors were placed. Here we expand the focus to the full flow, which naturally includes regions beyond the domain of dependence of the observations.  Figure~\ref{fig:errorfields} shows the r.m.s. errors in the instantaneous (a,b) pressure $\epsilon_{p}=\langle (p' - {p'}^{\!{\scriptscriptstyle{\top}}})^2 \rangle_z$ and (c,d) full state vector $\epsilon_{q}=\langle \left(\mathbf{q}-\mathbf{q}^{\scriptscriptstyle{\top}}\right)^\mathrm{T}\mathbf{M}\left(\mathbf{q}-\mathbf{q}^{\scriptscriptstyle{\top}}\right) \rangle_z$,
where $\mathbf{M}$ appropriately weights momentum, density, and internal energy (see equation \ref{eq:weightSolution}).  Near $2000 \leq \sqrt{Re}_x\leq 2400$, pressure deviations are amplifying in two regions: along the wall and then later ($\sqrt{Re_x}\gtrsim 2000$) near the boundary-layer edge $y\approx \delta_{99}$. The pressure deviation, seemingly disconnected in the two regions, appears to merge across the boundary layer to the wall by  $\sqrt{Re_x}\approx 2400$ and subsequently saturates beyond $\sqrt{Re_x}\gtrsim 2600$. Instantaneous differences in the turbulence have large errors in the pressure, but this error also appears to radiate above the turbulence. The visualization is consistent with errors in the radiated Mach waves~\citep{phillips1960generation}. Differences within the turbulence amount to differences in the radiated pressure as intense sound waves.
When the full state vector is considered, figure~\ref{fig:errorfields}~(c) shows that the deviation along the boundary layer edge is pronounced as early upstream as the inflow and more intense than at at the wall $y=0$.  This pattern can arise due to a missing inflow wave that EnVar could not identify or conversely one that was predicted by EnVar but which does not exist in the true flow; the latter is indeed the case as will be clarified by the inflow spectra. The absence of a similar error pattern in the pressure suggests that the erroneous wave is vortical in nature, supported near the boundary-layer edge.  
These errors advect downstream, amplify, and eventually spreads toward the wall. The impingement onto the wall takes place near the last three wall-pressure probes $2482 \leq \sqrt{Re_x}\ \leq 2567$, which coincides with the intense amplification of pressure errors in figure~\ref{fig:errorWallPressure}~(a). Notably, the visualization shows how inflow disturbances can elude detection from wall sensors until advection and diffusion spread their influence to the wall, potentially after participating in nonlinear interactions with other waves. {These results motivate future work on optimal sensors placements that maximize the sensitivity of the observations to the inflow condition and also assimilation of multiple sensing modalities in $\mathbf{m}$, e.g.\,synchronous Schlieren and PCB.} 


\begin{figure}
    \centering
    \includegraphics[width=1.\textwidth]{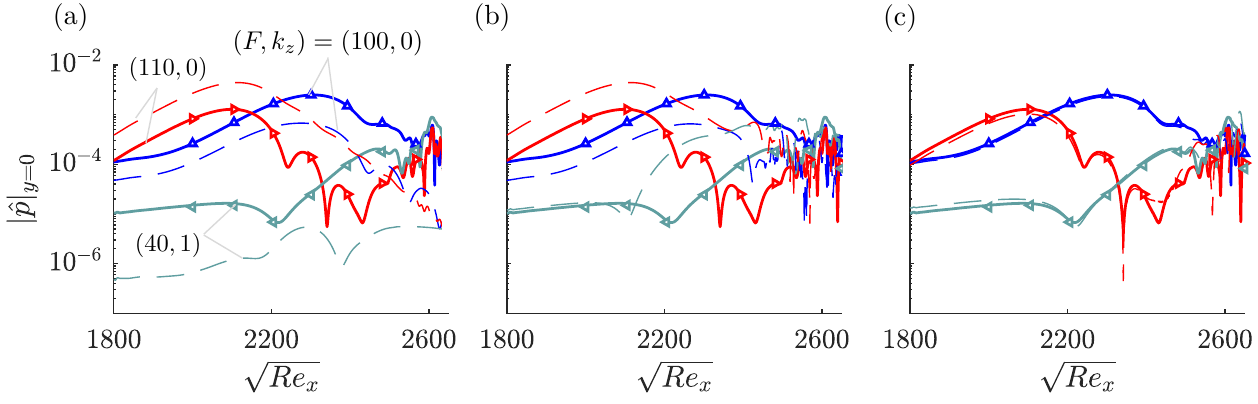}
    \caption{Streamwise development of wall-pressure spectra for (a)~E0, (b)~E1, and (c)~EnVar optimized framework at iteration $i=14$. (symbols) Observations; (solid) truth; (dashed) model prediction.}
    \label{fig:spectraInflowOnly}
\end{figure}

Since the flow is amenable to spectral analysis, we consider the two-dimensional Fourier transform in the span and in time.  We begin with the development of the three true inflow modes, only, and compare against predictions in figure~\ref{fig:spectraInflowOnly}.  The exponential amplification model adopted in cases E0 and E1 (panels a and b) mispredicts the inflow amplitudes of the second-mode instabilities $\modeFkz{100}{0}$ and $\mode{110}{0}$. Notably, their growth rates are similar to the truth up to $\sqrt{Re_x}< 2400$. While some modes are undergoing nonlinear interactions, the second mode evolves linearly to high amplitude (not shown), which is similar to observations in previous experiment-simulation comparisons of flow over a cone~\citep{kennedy2018investigation}. Without sufficient energy in $\modeFkz{40}{1}$ transition to turbulence is delayed beyond the simulation domain in~(a), {which underscores the importance of the three-dimensional waves that are often unavailable from experimental measurements and hence excluded from the analysis \citep{marineau2014mach,kennedy2018investigation,marineau2019analysis}. In case E1 (figure~\ref{fig:spectraInflowOnly}b),} the inflow amplitude of $\mode{40}{1}$ was accurately predicted by the exponential amplification model and its development is indistinguishable from the truth until $\sqrt{Re_x}\approx 2100$, after which the mode amplifies intensely and transition to turbulence is accelerated. The over-prediction of $\modeFkz{110}{0}$, together with $\mode{40}{1}$, spur nonlinearities earlier, specifically $\mode{70}{1}$, $\mode{30}{1}$, $\mode{10}{0}$, which feed back into $\mode{40}{1}$.  Only when the amplitudes of these three main instability waves are accurately predicted, relative to the true inflow condition, do their downstream nonlinear developments agree as in figure~\ref{fig:spectraInflowOnly}~(c).

\begin{figure}
    \centering
    \includegraphics[width=1.\textwidth]{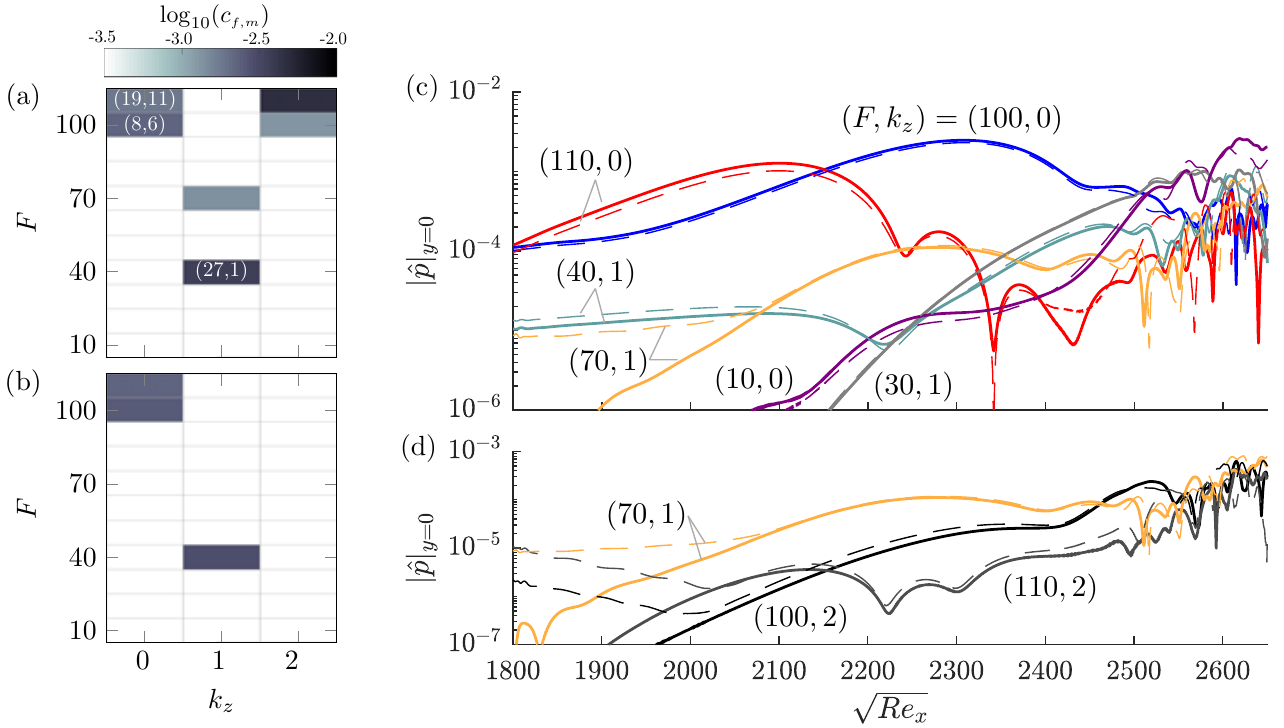}
    \caption{(a,b) Amplitudes of inflow instabilities: (a)~EnVar, (b)~truth. Marked values in (a) are the percent errors in the amplitude and phase relative to the true inflow.   (c,d) Streamwise development of the wall-pressure spectra: (solid) truth; (dashed) EnVar-predicted flow after 14 iterations.}
    \label{fig:spectraInflow}
\end{figure}

To summarize the outcome of the EnVar estimation using equally-weighted sensors, figure~\ref{fig:spectraInflow}~(a,b) show the amplitudes of the estimated and true inflow modes. Reported on the contours in (a) are percent errors in the amplitude and phase of the true inflow waves: $(\epsilon_A,\epsilon_\phi)=(19\%,11\%)$, $(8\%,6\%)$ and  $(27\%,1\%)$ for modes modes $\modeFkz{100}{0}$, $\mode{110}{0}$, and $\mode{40}{1}$, respectively. 
While these errors are rather large, they are important to discuss before attempting to reduce them further.
Here, we only note that we will successfully mitigate these errors in the following section to much lower levels, especially for second-mode instabilities\textemdash $(\epsilon_A,\epsilon_\phi)=(1\%,3\%)$, $(7\%,3\%)$ and  $(22\%,0.9\%)$\textemdash by an optimal weighting of observation stations. 
The large errors in the predicted inflow spectra highlight the difficulty of reconstructing the environmental disturbance from wall observations, even when the full nonlinear equations are adopted in the estimation. Even more, the results show that exceptional agreement of wall data and transition location may not translate into precise inflow interpretation. The core challenge is the nonlinear and chaotic nature of the system, for which multiple optima can confound interpretation and exacerbate the challenge of reducing inflow uncertainty. {In regard to these points, we emphasize that the transition mechanism is successfully reproduced, which is shown in figure~\ref{fig:spectraInflow}~(c) for the true inflow modes $\{(100,0),(110,0),(40,1)\}$. The figure also shows the evolution of three key waves $\{(70,1),(10,0),(30,1)\}$ that participate in important nonlinear interactions which culminate in breakdown to turbulence \citep[see][for detailed analysis of energy transfer among wave triads]{jahanbakhshi_zaki_2019}.
Figure~\ref{fig:spectraInflow}~(d) focuses on the inflow modes that are not part of the truth, including $\modeFkz{70}{1}$ which is reproduced from panel (c).  Notably, after a region of adjustment, all three erroneous inflow modes $\{(70,1), (100, 2), (110, 2)\}$ latch onto the trajectories of their counterparts in the true flow}. 
{This behaviour indicates that the observations are insensitive to the inflow amplitudes of these modes, which indeed do not compromise the accuracy of reproducing the observations or the fidelity of predicting the true transition mechanism\textemdash this point will be revisited in the following section.}

\section{Influence of sensor weighting on optimization}\label{sec:optimalWeighting}

In the previous section, the EnVar estimation of the unknown inflow reproduced the wall pressure observations and accurately predicted transition location and mechanism. Despite this success, the reduction in the cost function was relatively humble and the inflow spectra included appreciable errors. We contend that these symptoms are due to infusing all available observations with equal weights, some of which can negatively impact gradient-based optimization hence confounding interpretation. 
Observations can provide relevant physical information to discover unknown parameters; however, their information content may be duplicate and possibly yield conflicting interpretations. Observations may also be prone to uncertainty which affect optimization techniques.  These notions are all present in the mathematical formulation of the EnVar algorithm, which is influenced by the local shape of $\cost$ and relies on its local gradient and curvature.  
For example, chaotic dynamics such as in boundary-layer transition and turbulence imply that small changes in the control vector can yield large deviations in the observation matrix $\obsMat$, and hence appreciable reduction in the variance of subsequent ensembles; a remedy using variance inflation is thus needed.

From a geometric perspective, the chaotic dynamics also cause the surface of the cost functional to be highly oscillatory, or wrinkled.  Thus $\cost$ will have large local curvature restricting the optimizer in a shallow local optimum.   
The local curvature is described by the Hessian matrix, $\hessian$, which can be decomposed as $\hessian=\mathbf{Q}\Lambda \mathbf{Q}^{-1}$ where $\Lambda=\text{diag}(\lambda_1, \lambda_2,\ldots,\lambda_N)$ contains the eigenvalues and the columns of $\mathbf{Q}$ are the eigenvectors. We consider optimizing scalar invariants of the matrix $\Lambda$, e.g.\,condition number $\kappa=\lambda_\mathrm{max}/\lambda_\mathrm{min}$ where $\lambda_\mathrm{max}$ and $\lambda_\mathrm{min}$ are the maximum and minimum eigenvalues in $\Lambda$. To improve accuracy of inverting $\hessian$ or modify its curvature, we can manipulate the eigenspectrum \citep{ucinski2004optimal}.  For example, three potential strategies are to minimize $\kappa$ (K-optimality)~\citep{yan2018bayesian}, minimize $\lambda_{\textrm{max}}$ as will be adopted herein, maximize $\lambda_{\textrm{min}}$~(E-optimality)~\citep{colburn2011gradient,wilson2014trajectory}.

To adjust the eigenvalue spectrum of $\hessian$, we consider its two terms due to the observations ($\obsMat^\transpose \covMeasVec^{-1} \obsMat$) and the prior ($\pertMat^\transpose \covCntrVec^{-1} \pertMat$). In general, we have flexibility to adjust each term. For example, the measurement covariance $\covMeasVec$ can be made so large that all of the data are ignored and only the ensemble member control vectors affect the $\hessian$. In practice however, the measurement error $\covMeasVec$ is an outcome of irreducible signal noise and errors in the model that transforms the sensor signal into the observed flow quantity.  Therefore, increasing observation covariance is unjustified. Here we consider applying weights selectively to the observations, in both $\meas$ and $\obsMat$. With the applied weights, $\hessian$ can be reconstructed and its eigenvalue spectrum analysed. The weights are then adjusted until the eigenvalue spectrum has a desired property, e.g.\,minimizing the maximum eigenvalue of $\hessian$ thereby reducing extreme curvature. We emphasize that this scheme does not target flattening the cost functional in all control-vector directions; such strategy would lead to significant uncertainty in the solution of the optimization problem.

\begin{figure}
    \centering
    \includegraphics[width=0.99\textwidth]{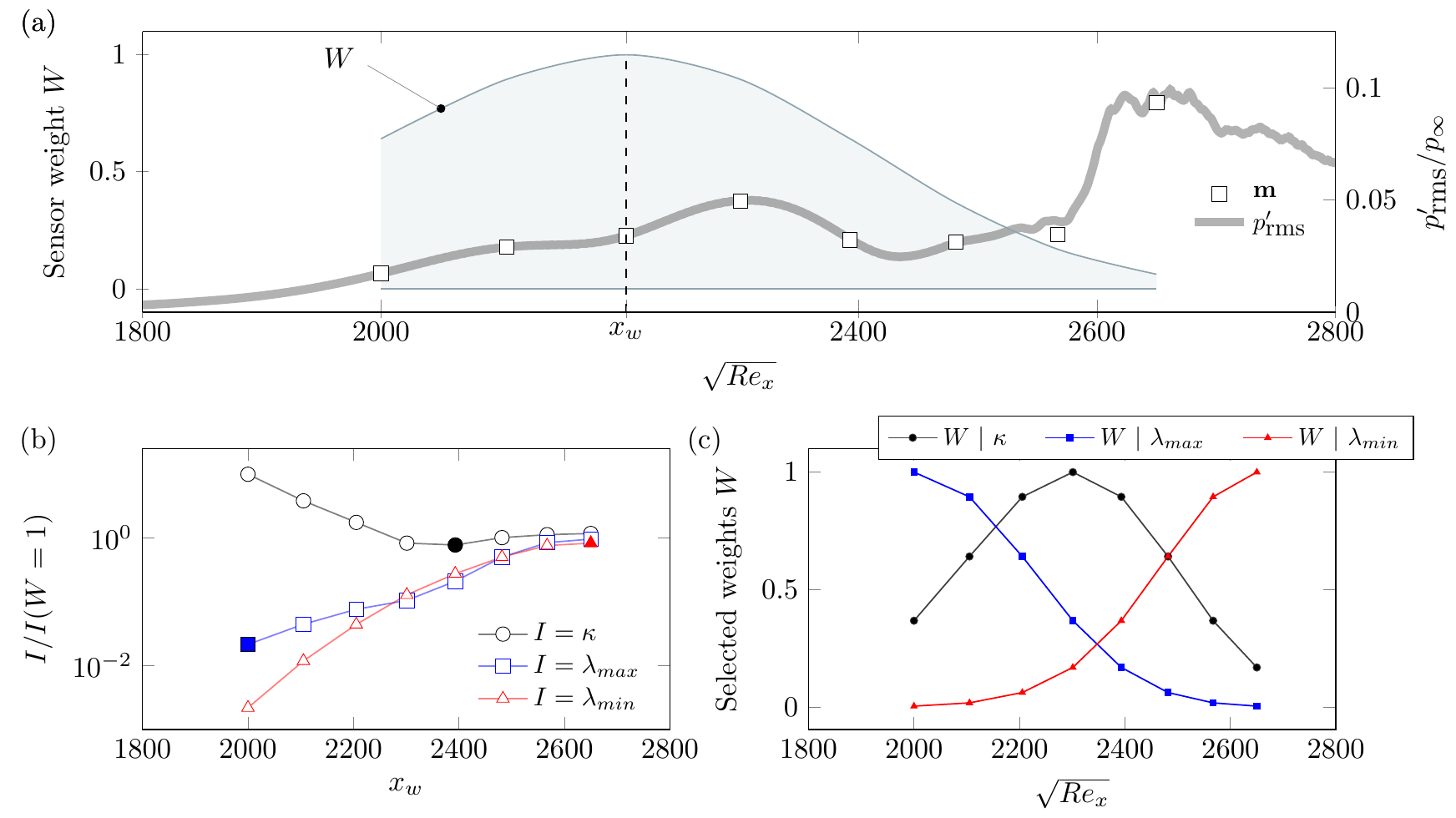}
    \caption{(a) Example of streamwise Gaussian weighting. (b) For each streamwise placement of the Gaussian weighting, $x_w = x_i$ for $i=\{1,2,\ldots,8\}$, the $\kappa$, $\lambda_{\mathrm{max}}$ and $\lambda_{\mathrm{min}}$ of $\hessian$ are reported, normalized by their values for uniform weighting. 
    Shaded symbols mark the optimal value for each criterion, and panel (c) shows the corresponding weighting.}
    \label{fig:weightingSchemesHypersonicBoundaryLayer}
\end{figure}

Although the expense of the DNS prohibits directly visualizing how the weighting affects the cost landscape, Appendix~\ref{app:lorenz} discusses a model Lorenz system where the landscape of $\cost$ is explicitly computed and the impact of various weighting strategies of observations are evaluated. Guided by those tests, we here examine weighting of wall-pressure observations based on their streamwise position as illustrated in figure~\ref{fig:weightingSchemesHypersonicBoundaryLayer}; spanwise-adjacent observations remain equally weighted since $z$ is a statistically homogeneous direction. We select a weighting function, 
\begin{equation}\label{eq:continuousWX}
    W(x)=\exp\left[\frac{-(x-x_w)^2}{\sigma_w^2}\right],
\end{equation}
where $x_w$ emphasizes a particular observation station and $\sigma_w$ determines the influence of neighboring sensors. The weighting (\ref{eq:continuousWX}) is embedded in the EnVar framework using a diagonal matrix $\Weight$ of size $N_d\times N_d$ where $N_d=N_t N_{x}^s N_{z}^s$ with elements $w_{l,l}=W(\mathbf{x}^s_{i,j})$ where $l=n + (i-1)N_t + (i-1)(j-1)N_t N_x^s$ and $n\in [1,N_t]$, $i\in[1,N_x^s]$, and $j\in[1,N_z^s]$.
The cost function, optimal weights, and Hessian are revised to
\begin{align}\label{eq:costLinearizationRev}
    \breve{\cost} &= \tfrac{1}{2}\norm{\Weight\meas - \Weight\projCntrVecMean - \Weight\obsMat \weights}^2_{\covMeasVec^{-1}} + \tfrac{1}{2}\norm{\pertMat\weights}^2_{\covCntrVec^{-1}},\\
    \breve{\weights} &= -\left(\obsMat^\transpose \Weight^\transpose\covMeasVec^{-1}\Weight \obsMat 
    + \pertMat^\transpose \covCntrVec^{-1}  \pertMat \right)^{-1}
    \left( \obsMat^\transpose \Weight^\transpose\covMeasVec^{-1} \left\{\Weight\projCntrVecMean  - \Weight\meas\right\}  \right)~\textrm{and}~\\
    \breve{\hessian} &= \left(\obsMat^\transpose \Weight^\transpose\covMeasVec^{-1}\Weight \obsMat 
    + \pertMat^\transpose \covCntrVec^{-1} \pertMat \right),
\end{align}
where the $(\breve{~~})$ indicates the influence of weighting $\Weight$. 

In practice, starting with members of the first ensemble, which are distributed about the PSE estimate $\cntrVec_o$, their evolutions are computed using DNS. 
The associated observation matrix was evaluated, and the Hessian $\hessian$ was constructed using various weighting strategies.  
The optimal weight matrix $\Weight$ that was identified based on the initial ensemble was adopted for all subsequent EnVar iterations~($i>0$). 
Dynamic updates to $\Weight$ as the observation and prior terms of the Hessian evolve during the nonlinear optimization is ongoing research, and may further benefit navigating the landscape of $\cost$.

The influence of the weighting on the eigenspectrum of the Hessian $\hessian$ is reported in figure~\ref{fig:weightingSchemesHypersonicBoundaryLayer}~(b). 
As one may expect, favourably weighting upstream sensors relative to those in the turbulent region reduces $\lambda_{\textrm{max}}$, or largest curvature, of $\hessian$.  
Similar to the simple transitional Lorenz model tested in detail in Appendix \ref{app:lorenz}, K- and E-optimality for observing the wall pressure in transitional boundary layers yield weightings that favour downstream sensors; these two approaches do not improve convergence of the state estimation algorithm because favouring sensors in chaotic regions create highly oscillatory cost functionals. Due to the computational expense of DNS within each EnVar iteration, we only consider the weighting that reduced $\lambda_{\textrm{max}}$, because the associated reduction in extreme curvature of the cost function correlated with improved performance in the Lorenz system.

The adjustments to the $\hessian$ not only impact the shape of the cost, but they also influence the subsequent distribution of ensemble members, which in turn affects convergence. For example, large curvature would require that ensemble members be sampled closer together in order to accurately approximate the local cost landscape; such tight distribution of members implies higher confidence in an uncertain control vector. \citet{wilson2014trajectory} made use of this last point by choosing experimental trajectories which maximize $\lambda_{\textrm{min}}$; the  minimum curvature in the $\hessian$ increased and the largest uncertainty of the control vector was reduced since $\covCntrVec \propto \hessian^{-1}$. However, the effect of chaotic behavior was not an apparent threat in their optimization. For the present transitional boundary layer configuration, chaos has an excessive impact on local curvature through $\hessian$, hindering optimization, which led us to consider the alternative approach of weighing sensor data which minimize $\lambda_{\textrm{max}}$. In addition to smoothing part of the cost landscape, the approach avoids generating an extremely tight distribution of ensemble members that would undesirably hamper exploration of the cost landscape.


 \begin{figure}
    \centering
    \includegraphics[width=0.8\textwidth]{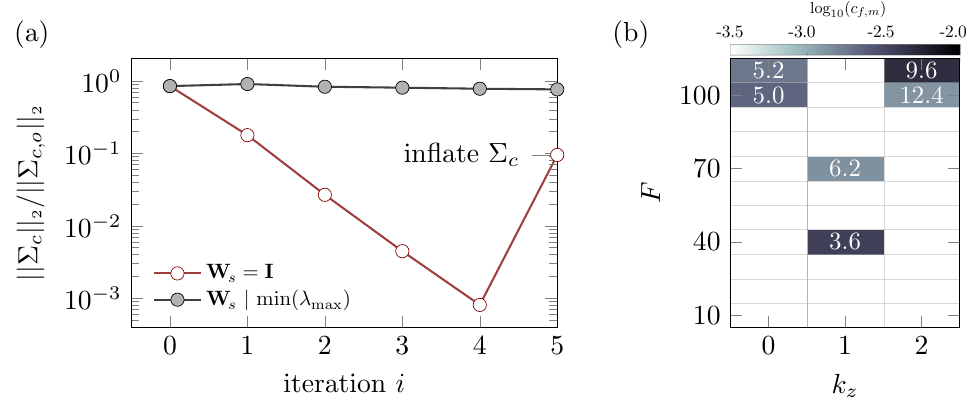}
    \caption{(a) Progression of the 2-norm of $\covCntrVec$ with EnVar iteration, $i\leq5$ and (b) amplitude of key modes. Marked values in (b) are the relative uncertainty $(\sigma_{f,m} / c_{f,m}, \sigma_{f,m} / \pi)\times 100$ for the amplitude and phases.}
    \label{fig:covariance}
\end{figure}
 
\FloatBarrier

The EnVar estimation procedure was repeated with the observations weighted using the matrix $\Weight$ that minimizes $\lambda_{\max}$ (see figure \ref{fig:weightingSchemesHypersonicBoundaryLayer}$c$). The  2-norm (square root of the maximum eigenvalue) of the control-vector covariance matrix is reported in figure~\ref{fig:covariance}~(a), normalized by its initial value in the EnVar iterations; this quantity is a proxy for the cost function curvature because the control-vector covariance is proportional to $\hessian^{-1}$.  The trend confirms that a weighting strategy that favours upstream observations guards against extreme curvature in $\hessian$ which would lead to excessive reduction of $\covCntrVec$. Uniform weighting, on the other hand, shrinks $\covCntrVec$ by over three orders of magnitude, and hence $\cost$ is dominated by the prior $\cntrVec$ term and the observations $\mathbf{m}$ are relatively ignored which stalls convergence. {A closer look at the optimal weighting is provided in figure~\ref{fig:covariance}~(b) where we report the normalized variance of the amplitude of each inflow mode, $\sigma_{f,m}/c_{f,m}$~{(\%)}, overlaid on the spectra. The results demonstrate smaller variance in the true inflow modes $\{(100,0), (110, 0), (40, 1)\}$ since their amplitudes impact the observations and hence the cost function.  In contrast, larger variances in modes $\{(70,1), (100, 2), (110, 2)\}$, which are not part of the true inflow condition, signal the lack of trust in those predictions. As discussed in connection with figure \ref{fig:spectraInflow}, our observations are insensitive to those modes, and hence their uncertainty does not appreciably change the cost function nor does it impact the transition mechanism.}

\begin{figure}
    \centering
    \includegraphics[width=0.99\textwidth]{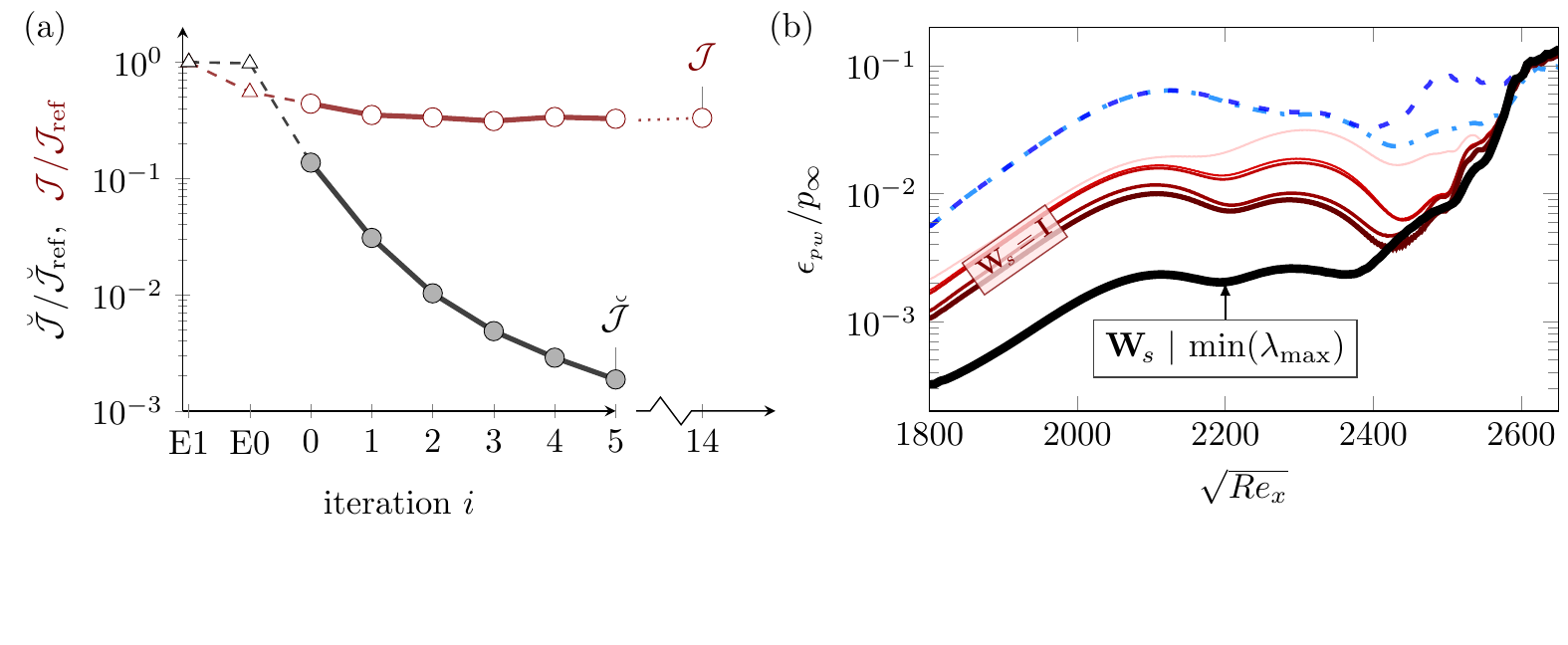}
    \vspace*{-24pt}\caption{(a) Comparison of the normalized cost function when observations are weighted to minimize the maximum eigenvalue of $\hessian$ versus uniform weighting. (b) Reduction of error in the wall pressure when observations are weighted (thick), relative to the uniform weighting results (thin lines reproduced from figure~\ref{fig:errorWallPressure}).}
    \label{fig:costWeighting}
\end{figure}

The behaviour of the cost function with EnVar iterations is reported in figure~\ref{fig:costWeighting}~(a), and shows two orders of magnitude better reduction than uniform sensor weighting. Cases E0 and E1 have approximately the same cost because the streamwise weighting reduces to $W\leq 0.17$ for $\sqrt{Re_x}\ge 2400$ figure~\ref{fig:weightingSchemesHypersonicBoundaryLayer}~(c), which is approximately the point where those flows either transition or remain laminar (see figure~\ref{fig:rms-cost}~(a)). 
Since error reduction in upstream sensors may not necessarily translate to a better overall prediction, we examine the error in the wall pressure as a function of downstream distance in in figure~\ref{fig:costWeighting}~(b).  The results show that accuracy is enhanced all the way upstream to the inflow and downstream to the onset of turbulence near $\sqrt{Re_x} \approx 2400$.

\begin{figure}
    \centering
    \includegraphics[width=0.99\textwidth]{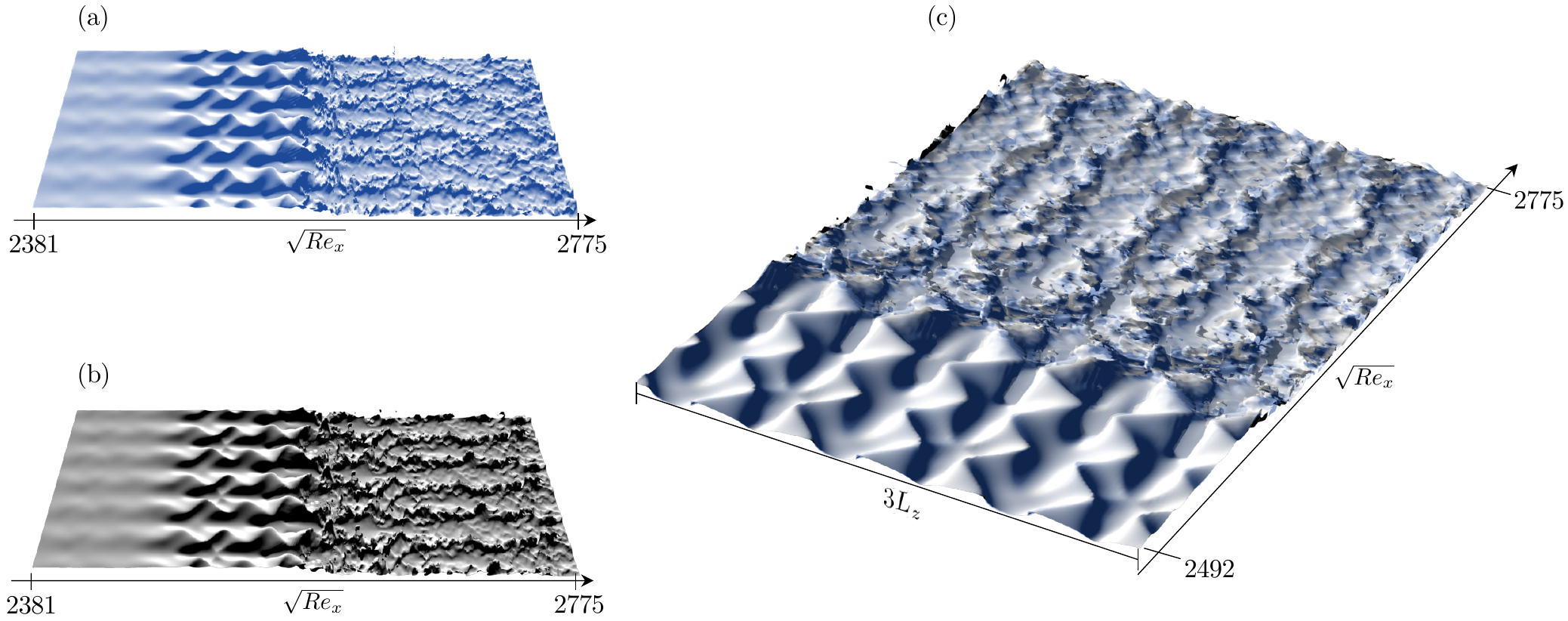}
    \vspace*{0pt}\caption{Iso-surface of instantaneous streamwise velocity $u=0.58$ at the end of the observation time horizon $t=T_a$, from (a) favourably weighted EnVar optimization ($i=5$) and (b) the true state. In (c), the two iso-surfaces are overlaid.}
    \label{fig:velocitycontour}
\end{figure}

Furthermore, the weighting of observations so as to minimize the maximum eigenvalue yields a better estimation of the inflow amplitudes and relative phases.  Figure~\ref{fig:spectraInflow}~(a) quantified the error for uniform weighting $(\epsilon_A,\epsilon_\phi)=(19\%,11\%)$, $(8\%,6\%)$ and  $(27\%,1\%)$ in predictions of the true modes $\modeFkz{100}{0}$, $\mode{110}{0}$, and $\mode{40}{1}$, respectively.  The new results with weighted observations reduce these errors to  $(\epsilon_A,\epsilon_\phi)=(1\%,3\%)$, $(7\%,3\%)$ and  $(22\%,0.9\%)$. 
The remaining discrepancies relative to the true inflow do not manifest in the sensor data. Most importantly, despite the difference between the estimated and true inflow, the downstream evolution of the two sets of modal amplitudes track one another downstream and the EnVar prediction captures the waves spawned by nonlinearity in the true flow.  Figure~\ref{fig:velocitycontour}~(a) shows the development of a streamwise velocity from the EnVar-predicted inflow with the weighted sensor signals, compared to the truth in panel (b), and then overlaid together in (c). Transition is accurately captured, both in terms of its onset location and mechanism.

\section{Discussion and conclusions}\label{sec:summary}
An ensemble-variational (EnVar) state estimation algorithm was described for infusing sensor data, or observations, in simulations of high-speed flows.   
The presentation was focused on transitional boundary layers, and the ability to interpret scarce observations such as wall-pressure data at discrete sensors; however, the methodology is general and applicable with any available observations and forward model.  
We sought to predict the inflow instability waves that reproduce the sensor data and, in doing so, predict the full flow state encoded in these observations. 
The first estimate of the unknown inflow condition was obtained using the linear parabolized stability equations, which is an effective approach when sensors are placed in regions where the perturbations dynamics are linear. In flight, the transition location is unknown due to the uncertainty in the energy of environmental disturbances and their interactions. Depending on the placement of sensors, they may be exclusively exposed to turbulence, which would violate PSE assumptions.  Therefore, subsequent iterations of the EnVar algorithm are performed using direct numerical simulations (DNS), although any forward nonlinear model can be considered.

The EnVar algorithm was adopted for the interpretation of wall-pressure observations at sixty-four sensing locations, from a $M=4.5$ flat-plate transitional boundary layer.
The sensors spanned the linear, nonlinear, and chaotic flow regimes. 
And the estimated flow successfully identified an inflow spectra that yielded accurate predictions of the wall pressure, transition location, and the transition mechanism. 
However, the reduction of the cost function was not sufficiently satisfactory, even with inflation of the ensemble covariance.  

Favourably weighting upstream observations reduces the maximum eigenvalue of the Hessian of the cost function, and hence the most extreme curvature of the cost function.  The outcome is improved convergence of the EnVar algorithm, in particular a more accurate prediction of the inflow disturbances and hence the flow.  This sensor weighting is counter to traditional strategies that attempt to minimize the condition number of the Hessian in order to ensure its accurate inversion, and to strategies that increase the Hessian minimum eigenvalue to improve trust in the predicted minimum.  The best weighting strategy is different precisely because of the present configuration which involves transition to chaotic flow.  The underlying cost functional landscape is tortuous and oscillatory, which hampers the optimization and leads to narrow search regions. Our weighting approach lessens that difficulty, and leads to better convergence of the algorithm.  Note that we achieve this goal without flattening the cost functional in all parameter directions, which would also be undesirable because an entirely flat cost functional would lead to significant prediction uncertainties.  Future work will apply the framework to experimental measurements, and will quantify the degree that measurement errors impact optimal sensor weighting, modality, and placement.

\appendix

\section{EnVar gradient accuracy}\label{app:accuracy}
\begin{figure}
    \centering
    \includegraphics[width=0.99\textwidth]{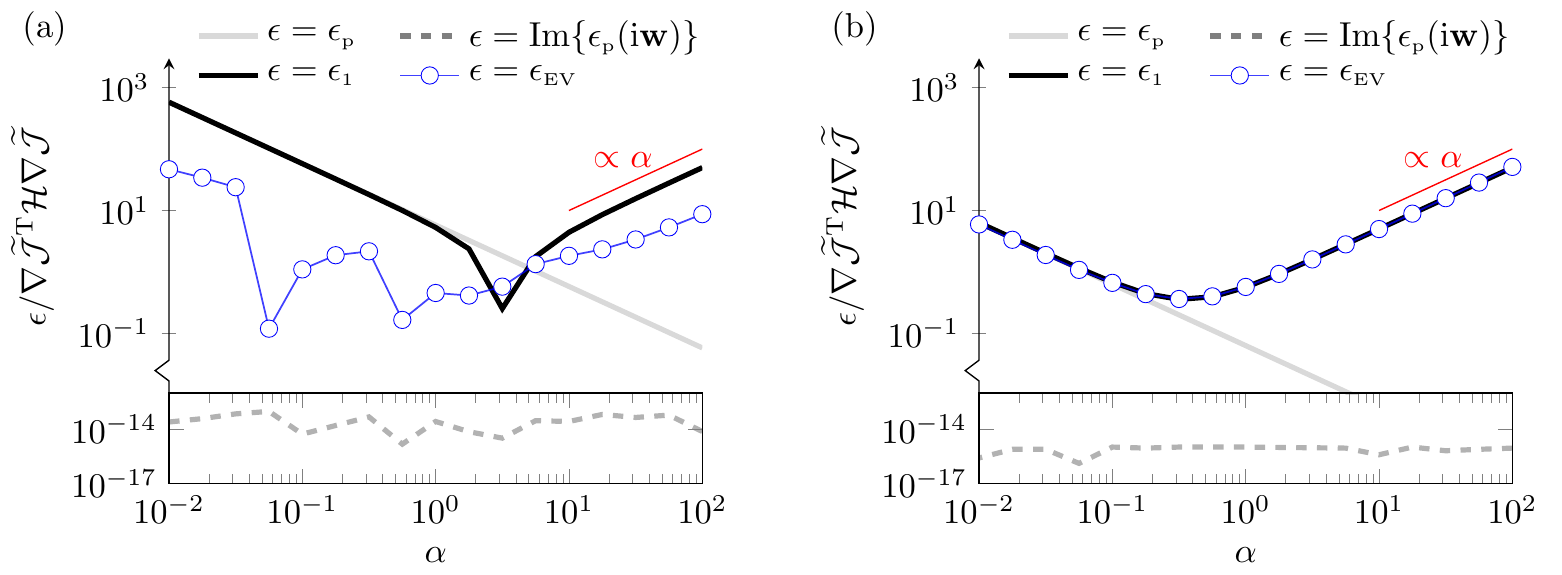}
    \caption{Accuracy of the cost-functional gradient computed using the EnVar approach in the transitional boundary layer: (a) Uniform weighting of observations $\Weight=\identity$ (EnVar iteration $i=14$);  (b) weighted observations $\Weight~|~\lambda_{\mathrm{max}}$ (iteration $i=5$).}
    \label{fig:gradaccuracy}
\end{figure}

The purpose of this appendix is to compare the gradient of the cost functional evaluated using EnVar to two finite-difference approximations, the first using the quadratic form of the cost functional $\LinCost$ and the second using the full nonlinear form $\cost$. 
As in similar assessments of gradient accuracy~\citep{vishnampet2015practical,buchta2016discrete}, we consider the Taylor series expansion of the quadratic cost function which is exact, 
\begin{equation}\label{eq:TaylorError}
    \LinCost(\mathbf{w}) =\LinCost(0) + \frac{\partial \LinCost}{\partial \mathbf{w}}^{\mathrm{T}} \mathbf{w} + \frac{1}{2}\mathbf{w}^\mathrm{T}\frac{\partial^2 \LinCost}{\partial \mathbf{w}^2}\mathbf{w}. 
\end{equation}
Consider the weights in the EnVar gradient direction, 
\begin{equation}
    \mathbf{w} =-\alpha \frac{\partial^2 \LinCost}{\partial \mathbf{w}^2}^{-1}\frac{\partial \LinCost}{\partial\mathbf{w}},
\end{equation}
substitute in (\ref{eq:TaylorError}), and rearranging yields an error measure,
\begin{equation}\label{eq:gradientError}
    \epsilon_{\scriptscriptstyle{\mathrm{p}}}\equiv\bigg{|}\underbrace{\frac{\LinCost(\mathbf{w})-\LinCost(0)}{\alpha} +\frac{\partial \LinCost}{\partial \mathbf{w}}^{\mathrm{T}}\frac{\partial^2 \LinCost}{\partial \mathbf{w}^2}^{-1} \frac{\partial \LinCost}{\partial \mathbf{w}}}_{\epsilon_{\scriptscriptstyle{1}}} -{\frac{\alpha}{2} \frac{\partial \LinCost}{\partial \mathbf{w}}^{\mathrm{T}}\frac{\partial^2 \LinCost}{\partial \mathbf{w}^2}^{-1} \frac{\partial \LinCost}{\partial \mathbf{w}}} \bigg{|}. 
\end{equation}
%
Equation (\ref{eq:gradientError}) should be identically zero. Any errors in $\epsilon_{\scriptscriptstyle{\mathrm{p}}}$ are due to machine round off, which will increase as we reduce $\alpha$. In addition, $\epsilon_{\scriptscriptstyle{\mathrm{1}}}$ is the error in approximating the derivative of a quadratic function, which is exact in EnVar, by a first-order finite-difference approximation.  This error will reduce with $\alpha$ until it is dominated by finite-precision effects.
Figure~\ref{fig:gradaccuracy} reports both errors, $\epsilon_{\scriptscriptstyle{\mathrm{p}}}$~(light curve) and $\epsilon_{\scriptscriptstyle{\mathrm{1}}}$~(dark curve). 
As anticipated, the $\epsilon_{\scriptscriptstyle{\mathrm{p}}}$ is only susceptible to finite precision.
This assertion is further confirmed by re-computing $\epsilon_{\scriptscriptstyle{\mathrm{p}}}$ using a complex step $\weights\rightarrow \imag\weights$ in (\ref{eq:gradientError}).
The imaginary part of the error yields
\begin{equation}\label{eq:gradientErrorComplexStep}
    \mathrm{Im}\left\{\epsilon_{\scriptscriptstyle{\mathrm{p}}}(\imag\mathbf{w})\right\}=\bigg{|}\mathrm{Im}\left(\frac{\LinCost(\imag\mathbf{w})}{\alpha}\right) +\frac{\partial \LinCost}{\partial \mathbf{w}}^{\mathrm{T}}\frac{\partial^2 \LinCost}{\partial \mathbf{w}^2}^{-1} \frac{\partial \LinCost}{\partial \mathbf{w}}\bigg{|} =0,
\end{equation}
which bypasses the effect of differencing in (\ref{eq:gradientError}.  As a result, the errors which are reported in figure~\ref{fig:gradaccuracy} (dashed curve) decrease to insignificant values. 

The above approaches verify the gradient implementation for a quadratic cost functional, $\LinCost$.
For a cost function that is not strictly quadratic $\cost$, its local tangent direction may not correspond with the EnVar-based gradient. 
The deviation is evaluated by comparing a finite-difference approximation of the gradient to an EnVar approximation, 
\setlength\fboxsep{1pt}
\begin{equation}
    \epsilon_{\scriptscriptstyle{\mathrm{EV}}}=\bigg{|}\frac{\boxed{\cost(\mathbf{w})}-\LinCost(0)}{\alpha} +\frac{\partial \LinCost}{\partial \mathbf{w}}^{\mathrm{T}}\frac{\partial^2 \LinCost}{\partial \mathbf{w}^2}^{-1} \frac{\partial \LinCost}{\partial \mathbf{w}}\bigg{|}.
\end{equation}

The gradient accuracy is reported in figure~\ref{fig:gradaccuracy} for two cases: (a) when the sensors are equally weighted and (b) when the upstream ones are favoured.  
In the former case, the degree of accuracy is affected by nonlinearity of the forward system. 
We find that the EnVar approximation is suitable for an appropriate choice of the step size. 
The results further verify the EnVar gradient approximation is accurate, and it improves for favourable observation weighting.  

\section{Estimate of control-vector covariance}\label{app:matrixIdentity}

In this appendix, we motivate the relationship between the Hessian of $\LinCost$ and the control-vector covariance matrix $\covCntrVec$ which is used to generate ensemble members. 
We consider a stationary point $\nabla_\mathbf{w} \LinCost=0$, and evaluate the total variation of the cost-function gradient,
\begin{equation}
     \delta \left(\frac{ \partial\LinCost}{\partial \mathbf{w}} \right) =  \frac{\partial^2 \LinCost}{\partial \weights^2} \delta \weights + \frac{\partial^2 \LinCost}{\partial \meas\partial \weights} \delta \meas = 0.
     \label{eq:variationlogLik}
 \end{equation}
The variation of the weights due to perturbations in observations is therefore, 
 \begin{equation}\label{eq:measuAmplifiesWeights}
     \delta \weights = -\frac{\partial^2 \LinCost}{\partial \weights^2}^{-1}\frac{\partial^2 \LinCost}{\partial \meas\partial \weights} \delta \meas, 
 \end{equation}
where the second derivatives can be evaluated from the quadratic cost function~(\ref{eq:costLinearization}), 
\begin{align}\label{eq:secondDerivatives}
      \frac{\partial^2 \LinCost}{\partial \weights^2}=\obsMat^{\transpose} \covMeasVec^{-1}  \obsMat + \pertMat^{\transpose} \covCntrVec^{-1}\pertMat~~~~\textrm{and}~~~~
      \frac{\partial^2 \LinCost}{\partial \meas\partial \weights} = -\obsMat^{\transpose}\covMeasVec^{-1}.
\end{align}
Then, the covariance of weights is
\begin{align}
\expectation\left[\delta \weights \delta \weights^{\transpose}\right]
     &=\nonumber \\
     &\expectation\left[\left(\obsMat^{\transpose} \covMeasVec^{-1}  \obsMat + \pertMat^\transpose \covCntrVec^{-1} \pertMat\right)^{-1} \obsMat^{\transpose} \covMeasVec^{-1}  \obsMat \left(\obsMat^{\transpose} \covMeasVec^{-1}  \obsMat + \pertMat^\transpose \covCntrVec^{-1} \pertMat\right)^{-1} \right],\label{eq:covWeightsFullwithCov}
 \end{align}
 which can be simplified to
 \begin{equation}
     \underbrace{\expectation\left[\delta \weights \delta \weights^{\transpose}\right]}_{\Sigma_\weights} \approx \bigg(\underbrace{\obsMat^{\transpose} \covMeasVec^{-1}  \obsMat + \pertMat^\transpose \covCntrVec^{-1} \pertMat}_{\hessian}\bigg)^{-1}. \label{eq:fullcovariance}
 \end{equation}
The left side of (\ref{eq:fullcovariance}) is the covariance of weights and the right side is the inverse Hessian of $\LinCost$. This approximation holds for $\norm{ (\obsMat^\transpose\covMeasVec^{-1}\obsMat)^{-1}(\pertMat^\transpose \covCntrVec^{-1} \pertMat)}\ll \norm{\identity}$, which is not immediately clear by inspection of (\ref{eq:covWeightsFullwithCov}).  However, we clarify the simplification of (\ref{eq:covWeightsFullwithCov}) to (\ref{eq:fullcovariance}) by recognizing that the right side of (\ref{eq:covWeightsFullwithCov}) has the form $(\AMat+\BMat)^{-1} \AMat (\AMat+\BMat)^{-1}$ where $\mathbf{A}$ and $\mathbf{B}$ are same-sized invertible matrices. The expression is simplified in the following steps:
 \vspace{-6pt}
 \begin{align}\nonumber
    (\AMat+\BMat)^{-1} \AMat (\AMat+\BMat)^{-1}&=(\identity+\AMat^{-1} \BMat)^{-1}\AMat^{-1} \AMat (\identity+ \AMat^{-1} \BMat)^{-1}\AMat^{-1} \\ \nonumber
    &=\left[\AMat (\identity+ \AMat^{-1} \BMat) (\identity+ \AMat^{-1} \BMat) \right]^{-1}\\\nonumber
    &=\left[(\AMat+\BMat) (\identity+ \AMat^{-1} \BMat)\right]^{-1}\\\nonumber
    &\approx (\AMat+\BMat)^{-1}
 \end{align}
 provided $\norm{ \AMat^{-1} \BMat}\ll \norm{\identity}$.
 

Next, we show the relationship between the covariance of the weights and the control-vector covariance matrix. 
By its definition, the estimate of the control vector is a weighted superposition of perturbations to the prior, and hence its variation is $\delta \mathbf{c} = \mathbf{P} \delta\weights$. Then, the control-vector covariance matrix is 
\begin{align}\nonumber
    \underbrace{\expectation \left[\delta \cntrVec \delta \cntrVec^\transpose \right]}_{\covCntrVec^\star}&=
      \pertMat \expectation\left[\delta \weights \delta \weights^\transpose \right] \pertMat ^\transpose \\ \label{eq:covarianceC} &\approx  \pertMat \hessian^{-1} \pertMat^{\transpose},
\end{align} 
using (\ref{eq:fullcovariance}). 
By updating the ensemble members using ${\pertMat_{\scriptscriptstyle{{i+1}}} = \sqrt{\Nens-1}\pertMat_{\scriptscriptstyle{{i}}} \hessian^{-1/2} \mathbf{U}}$, where $\mathbf{U}$ is a random, mean preserving, unitary matrix, we guarantee that the control-vector covariance matrix, $\covCntrVec^{\scriptscriptstyle{{i+1}}}=1/(\Nens-1)\pertMat_{\scriptscriptstyle{{i+1}}} {\pertMat_{\scriptscriptstyle{{i+1}}}}^{\!\!\!\!\!\!\!\transpose}$, adheres to the relationship (\ref{eq:covarianceC}).


\section{Observation-infused simulations of Lorenz system}\label{app:lorenz}

In this appendix, we use a low-dimensional chaotic system to (i) probe performance of the EnVar technique described in \S\ref{sec:DAFormulation} and to (ii) contrast different sensor-weighting techniques.
The system models thermal convection in a two-dimensional domain that is periodic in the streamwise direction and forced by gravity ($g$) and a temperature difference between the upper and lower boundaries, $\Delta T=T(0)-T(H)$. 
The fluid is ideal with thermal diffusivity $k=\kappa/\rho c_p$, kinematic viscosity $\nu$ and coefficient of thermal expansion $\alpha$. 
The low-dimensional dynamics \citep{Saltzman1962,Lorenz1963} consider the temperature fluctuations ($T'$) and velocity potential $\psi$ in Fourier space, 
\begin{align}\label{eq:temperature}
    T'(x,y,t)&\propto Y(t) \sqrt{2} \cos\left(\tfrac{1}{2}k_a x \right)\sin\left(\tfrac{1}{2}k_y y \right)-Z(t)\sin\left(k_y y \right)\\\label{eq:velocitypotential}
    \psi(x,y,t)&\propto X(t)\sqrt{2} \sin\left(\tfrac{1}{2}k_a x \right)\sin\left(\tfrac{1}{2}k_y y\right),
\end{align}
where $k_a = a 2\pi/H$ and $k_y = 2\pi/H$ are the streamwise and wall-normal wavenumbers.  
The time dependent Fourier coefficients $\mathbf{X} = [X,Y,Z]^T$ are solutions to,
\begin{align}\label{eq:Lorenz}
    \frac{dX}{dt}=\sigma(Y-X),\quad
    \frac{dY}{dt}=-XZ+\rho X -Y,\quad\textrm{and}\quad 
    \frac{dZ}{dt}= XY - \beta Z, 
\end{align}
with initial perturbation to the system $[X,Y,Z](t=0)=[X_o, Y_o, Z_o]$. 
The specific configuration we consider follows~\cite{Saltzman1962}: $a=\sqrt{1/2}$, $\sigma=10$, $\rho=28$, and $\mathbf{X}_o=[0, 1,  0]^{\transpose}$. 

\begin{figure}
    \centering
    \includegraphics[width=0.99\textwidth]{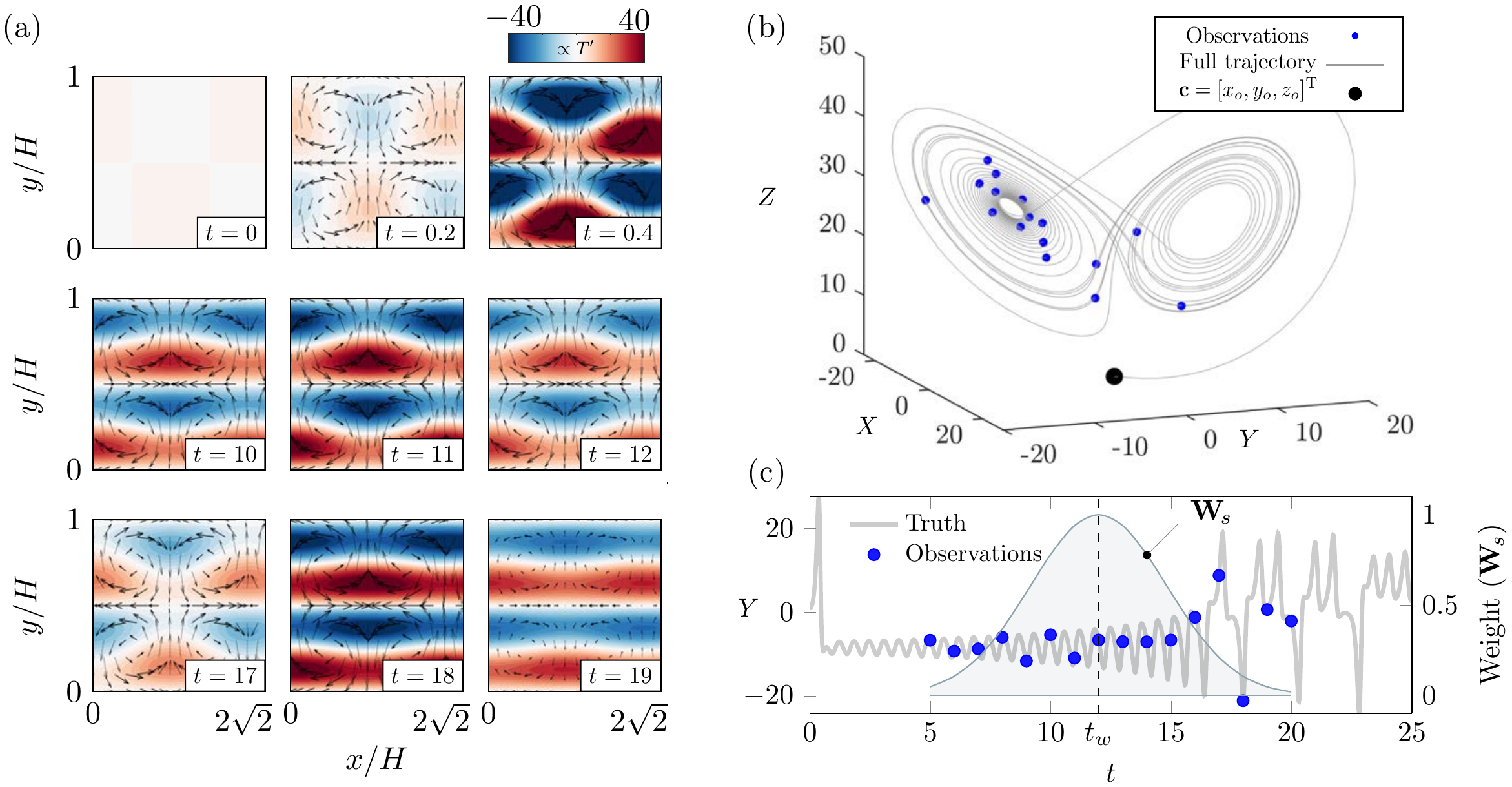}
    \caption{(a)~Instantaneous temperature (colour) and velocity magnitude (arrows). (b)~True phase-space trajectory (thick,gray), observation locations (small symbols), and initial condition~(large symbol). (c)~Time evolution of $Y$: observations (symbols), truth (thick,gray), and Gaussian weighting strategy in (blue/gray).}
    \label{fig:lorenzBehavior}
\end{figure}

Figure~\ref{fig:lorenzBehavior}~(a) shows transition to chaos. The initial disturbance is comprised only of temperature fluctuations, which subsequently amplify and generate velocity perturbations. Due to $\Delta T$, cold fluid cools further and conversely hot heats up, amplifying the instability by $t\approx 0.4$.  With gravity, the convection cells overturn. The cells oscillate about a new equilibrium state, which amplifies and overturns again by $t>15$. Beyond this time, the behavior is chaotic. The trajectory of the state $\mathbf{X}$ is shown in figure~\ref{fig:lorenzBehavior}~(b), and the observed quantity, $Y$, is shown in (c).  
The behaviour in figure~\ref{fig:lorenzBehavior}~(c) is similar to boundary-layer transition: amplification about an equilibrium (laminar) state and transition to chaos (turbulence). 

Of the three state variables, we observe part of  $Y$.  We then attempt to (i) identify the complete initial state $\mathbf{X}_o$ that (ii) reproduces observations, and (iii) reconstruct the solution trajectory beyond the observed horizon. 
Specifically, the observation vector is given by $\meas=\left[Y(5),Y(6), \ldots, Y(t_f)\right]^{\transpose}$ to time $t_f=20$.
The choice was motivated by the transitional boundary layer (figure~\ref{sec:simConfig}) where observations are available downstream of the inflow and across the nonlinear and chaotic regimes. 
The control vector we seek to identify is $\cntrVec=\left[X_o, Y_o, Z_o \right]^\transpose$. Since the observations $\meas$, early on ($t<15$), oscillate near $Y=-6\sqrt{2}$ in figure~\ref{fig:lorenzBehavior}~(c), we consider the initial perturbation originated near equilibrium point, $\mathbf{c}_o=[-6\sqrt{2},-6\sqrt{2},27]^\transpose$. The system (\ref{eq:Lorenz}) is integrated using standard fourth-order Runge--Kutta method and discretized with a fixed time step $\Delta t=0.01$, which is the same numerical method that generated $\meas$. The parameter and observation covariance matrices used in the EnVar method  are set to $\covCntrVec=\sigma_c\mathrm{I}$ and $\covMeasVec=\sigma_m\mathrm{I}$ where $\sigma_c=1$ and $\sigma_m=10^{-3}$, respectively.

\begin{figure}
    \centering
    \includegraphics[width=0.99\textwidth]{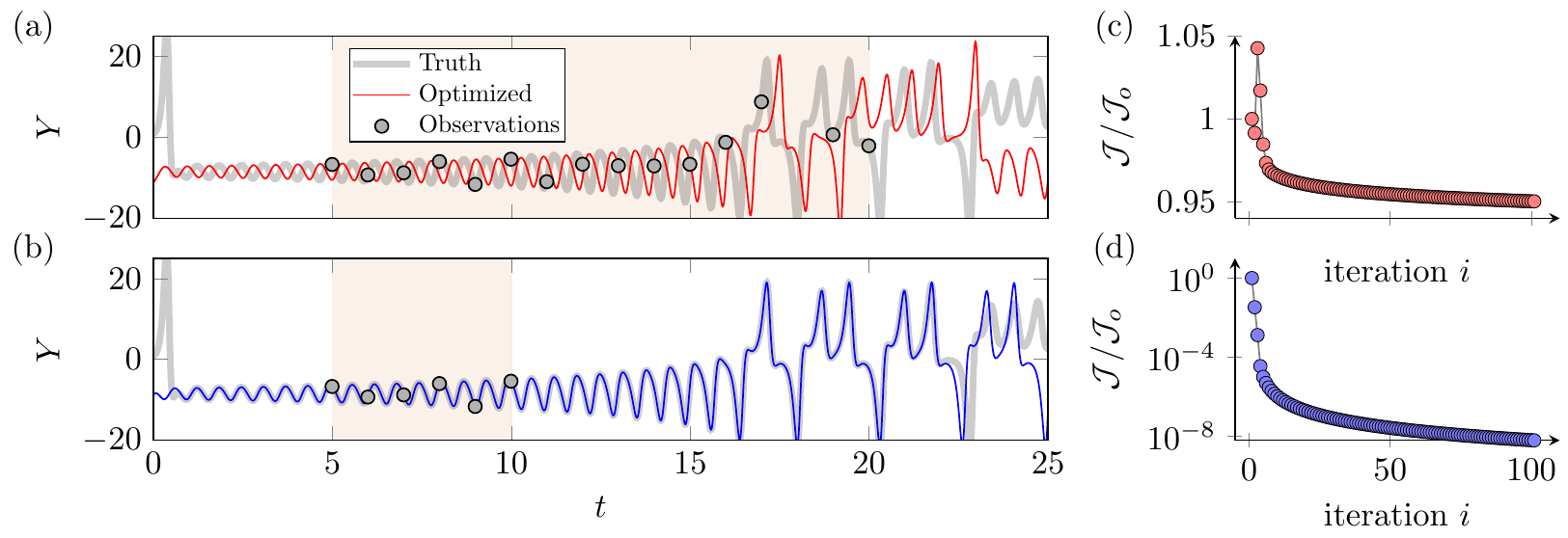}
    \caption{(a,b) The estimated time evolution of $Y$ after one hundred EnVar iterations. Shaded intervals identify the extents of the observation time horizons, $\meas=\left[Y(5),Y(6),\ldots,Y(t_f)\right]^{\transpose}$. (a) Target observation horizon $t_f=20$; (b) shorter test horizon $t_f=10$. (c,d) Convergence history of the normalized cost function.}
    \label{fig:lorenzEarlierMeasurements}
\end{figure}

Results from the EnVar initial-state estimation are reported in figure \ref{fig:lorenzEarlierMeasurements}, where two time horizons of observations are contrasted.  For each case, one hundred iterations were performed.
When the time horizon includes observations within the chaotic regime (figures \ref{fig:lorenzEarlierMeasurements}~(b)), the predictions are qualitatively similar to the results for the transitional boundary layer:  the time of transition in the thermal system near $t\approx 15$ appears to be captured, and a very modest reduction in cost is achieved in (c). 
When only the first five observations are considered (figures~\ref{fig:lorenzEarlierMeasurements}~(b)), the estimated initial condition reproduces the measurements with higher accuracy and predicts the evolution for longer duration.
In both cases, it is important to note that the predicted state deviates from the true initial condition. 
In addition, while disregarding sensor data in the second case may appear \textit{ad hoc}, this test indicates (i) that earlier acquisition provides better information about $t=0$ than later times, (ii) that later observations negatively impact gradient-based optimization, or (iii) that a mixture of these two effects takes place. The first proposition may appear natural, but is not necessarily accurate: small perturbations in $X_o$ or $Z_o$ may be entirely missed by early observations of $Y$.  In the same manner, a wall-pressure sensor near the inflow may be entirely insensitive to an inflow disturbance whose wall signature only develops downstream. 

\begin{figure}
    \centering
\includegraphics[width=0.95\textwidth]{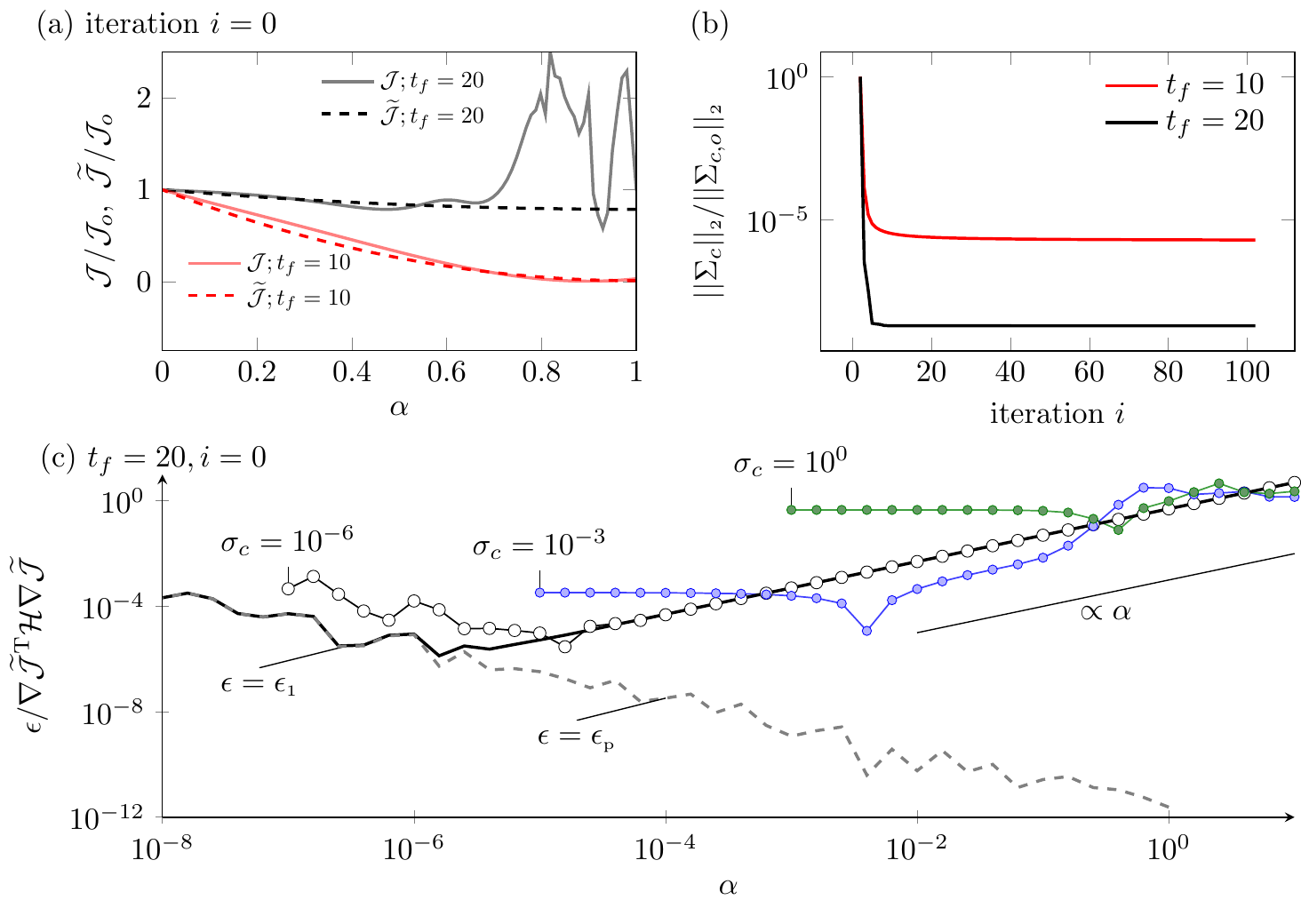}
    \caption{{(a) Effect of discarding observations on the shape of the cost function, evaluated in the gradient direction, at the initial EnVar iteration. Full observation horizon $t_f=20$; truncated horizon $t_f=10$. (b) The 2-norm of the covariance matrix of the control vector as a function of EnVar iterations. (c) Errors in the gradient of the cost functional evaluated using EnVar versus step size $\alpha$ in the gradient direction.  
    Symbols: errors relative to finite difference $\epsilon_{\scriptscriptstyle{\mathrm{EV}}}$, at various levels of ensemble covariance  $\covCntrVec=\sigma_c\identity$.  
    See Appendix \ref{app:accuracy} for definitions of $\epsilon_1$ and $\epsilon_p$.}}
    \label{fig:lorenzEarlierMeasurementsShape}
\end{figure}

In order to diagnose the convergence of $\cost$ in figures~\ref{fig:lorenzEarlierMeasurements}~(c,d), we probe the following: cost function shape along the gradient direction; the progress of $\covCntrVec$ which depends on the Hessian of $\LinCost$; and the gradient accuracy using equations (\ref{eq:gradientError}-\ref{eq:gradientErrorComplexStep}). Figure~\ref{fig:lorenzEarlierMeasurementsShape}~(a) shows the landscape of $\cost(\mathbf{w})$ and $\LinCost(\mathbf{w})$ as a function of the step size in the gradient direction 
\begin{equation}\nonumber
    \mathbf{w} =-\alpha \frac{\partial^2 \LinCost}{\partial \mathbf{w}^2}^{-1}\frac{\partial \LinCost}{\partial\mathbf{w}},
\end{equation}
with $0\leq \alpha \leq 1$, where $\alpha=1$ is the optimal step size for a quadratic function. 
When only early observations are considered, the approximation $\LinCost$ follows more closely $\cost$ which appears amenable to gradient-based optimization.  However, when later observations are included, $\cost$ is oscillatory and its linearization $\LinCost$ is only valid for small distances from the previous estimate of the control vector.  Notably in figure~\ref{fig:lorenzEarlierMeasurementsShape}~(b), $||\covCntrVec||_2$ is two orders of magnitude smaller in the latter case. As such, subsequent ensemble members will remain close to the prior $\cntrVec$, the observations $\meas$ are ignored, and the $\cost$ remains flat. 
Figure~\ref{fig:lorenzEarlierMeasurementsShape}~(c) shows that the initial gradient is relatively accurate. The first-order gradient accuracy can be improved by constructing ensemble members closer to the mean (e.g. $\sigma_c=10^{-3}$ and $\sigma_c=10^{-6}$). However, as the EnVar framework progresses, the posterior statistics of the Hessian reduce uncertainty of the control vector, thereby reducing $\covCntrVec$ as in~figure~\ref{fig:lorenzEarlierMeasurementsShape}~(b). 

The conclusion from figure~\ref{fig:lorenzEarlierMeasurementsShape} is that incorporating later observations, which are chaotic, yields an oscillatory cost functional shape and unpredictable changes in the cost, even though the gradient near $\alpha=0$ is well approximated.  Cost reduction is subsequently hindered because of extreme reduction in the control-vector covariance due to large sensitivity (large curvature in $\hessian$) from observations that are very sensitive to the initial condition. While not shown, control-vector covariance inflation alone does not improve the cost reduction when later observations are considered.

\begin{figure}
    \centering
    \includegraphics[width=1\textwidth]{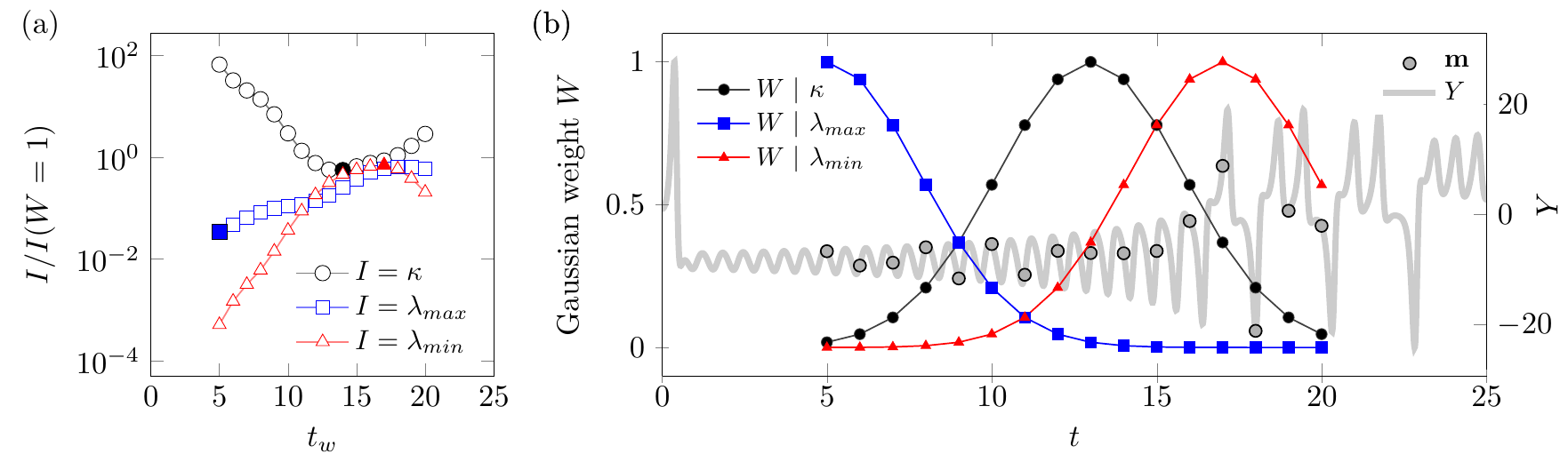}

    \caption{
    (a) For each temporal placement of the Gaussian weighting, $t_w = t_i$ for $i=[5,6,\ldots,20]$, the $\kappa$, $\lambda_{\textrm{max}}$ and $\lambda_{\textrm{min}}$ of $\hessian$ are reported, normalized by their values for uniform weighting. Shaded symbols mark the optimal value for each criterion, and panel (b) shows the corresponding weighting.}
    \label{fig:costLorenz}
\end{figure}

Instead of simply eliminating observations, sensor weighting strategies are evaluated for the Lorenz system. 
We adopt a weighting function, 
\begin{equation}\label{eq:continuousW}
    W=\exp\left[\frac{-(t-t_w)^2}{\sigma_w^2} \right],
\end{equation}
which biases a particular observation time $t_w$, and $\sigma_w$ controls the influence of neighboring sensors.  
The impact of the weights on the performance of the EnVar algorithm can be related to the eigenspectrum behaviour of $\hessian$, specifically the following: (i) Minimizing the condition number $\kappa$ promotes its accurate inversion and hence accurate evaluation of the ensemble weights. (ii) Maximizing the minimum eigenvalue $\lambda_{\textrm{min}}$ reduces the largest uncertainty in the estimated control vector \citep{ucinski2004optimal,wilson2014trajectory,colburn2011gradient}.  
(iii) Minimizing the maximum eigenvalue $\lambda_{\textrm{max}}$ guards against the most extreme curvature, which we also interpret as the most egregious over-estimate of trust in a component of the control vector. 
Figure~\ref{fig:costLorenz}~(a) reports the impact of different $t_w$ on these invariants of $\hessian$, and panel (b) shows the Gaussian profile that best accomplishes each strategy.  The results demonstrate that $\kappa$ is minimized by favourably weighting the intermediate observations; maximizing $\lambda_{\textrm{min}}$ is achieved by favouring later ones; minimizing $\lambda_{\textrm{max}}$ places the focus on early observations.  These trends in the context of chaotic convection apply similarly to transitional boundary layers when observations span the linear, nonlinear, and chaotic regimes.

\begin{figure}
    \centering
    \includegraphics[width=0.99\textwidth]{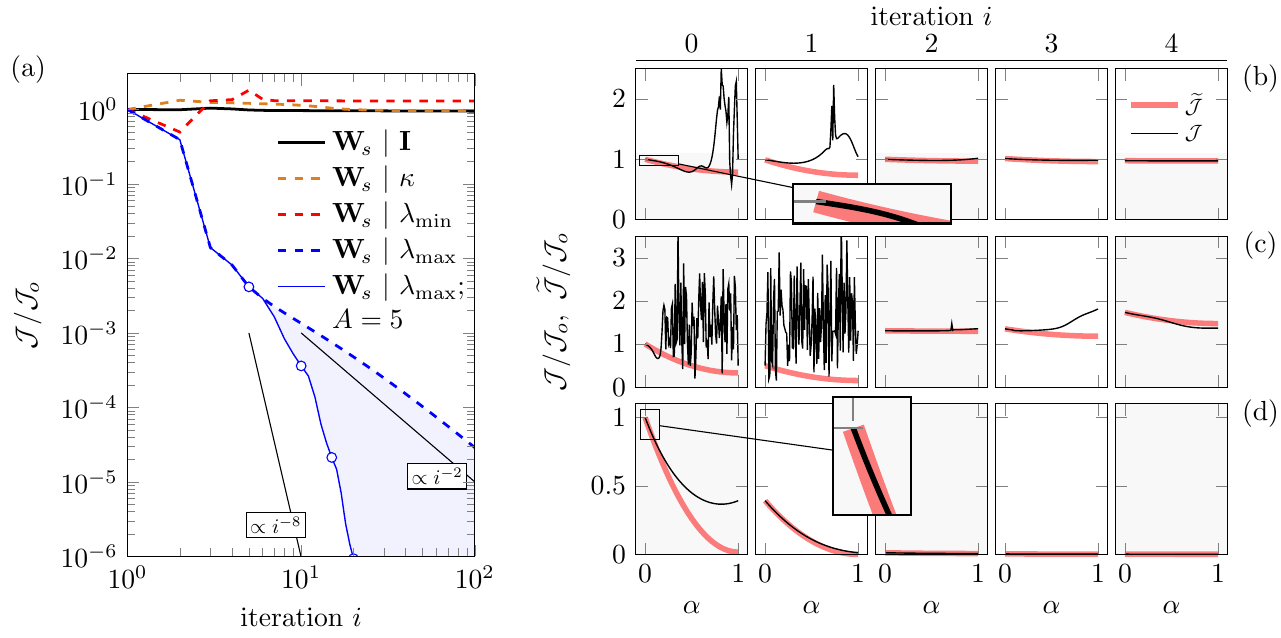}
    \caption{(a) Comparison of normalized cost functional during EnVar optimizations that adopt different sensor weighting ($\Weight$) schemes. (b-d) Cost functional in the gradient direction versus the step size,  for the first five iterations. (b) $\Weight=\identity$; (c) $\Weight~|~\lambda_\mathrm{min}$; (d) $\Weight~|~\lambda_\mathrm{max}$.}
    \label{fig:detailedCostLorenz}
\end{figure}

The performance of the EnVar algorithm with the above sensor-weighting strategies is reported in figure~\ref{fig:detailedCostLorenz}~(a). The best reduction of the cost function is achieved by favouring the earlier observations, or minimizing $\lambda_{\textrm{max}}$ thus guarding against extreme curvature. This trend is consistent with the earlier test in figure~\ref{fig:lorenzEarlierMeasurements}~(c,d). The figure also shows that convergence is further accelerated by applying covariance inflation every five iterations.  Previous research has proposed amplifying the parameter covariance in order to avoid over-estimating confidence in them \citep[e.g.][]{anderson1999monte,colburn2011gradient,da2018ensemble}. In doing so, convergence accelerates by several orders of magnitude more.

Ultimately, weighting of observations aims to modify the landscape of the cost function.  Figures~\ref{fig:detailedCostLorenz}~(b-d) display the landscape during the first five EnVar iterations, using the same approach discussed for figure~\ref{fig:lorenzEarlierMeasurementsShape}~(a). Deviation between the thin and thick curves in each panel is due to the linear approximation within $\cost$ by $\LinCost$, or precisely full nonlinear observations $\mathcal{H}(\cntrVec+\pertMat\weights)$~(thin) versus the linearized $\mathcal{H}(\cntrVec)+\obsMat\weights$ (thick). 
The cost function $\cost$ is oscillatory when weighting all observations equally or favouring later ones. Oscillations are most pronounced for the latter case which aims to increase $\lambda_{\textrm{min}}$, or reduce the largest control-vector variance\textemdash a strategy that is inadequate for a chaotic cost function and where late observations are extremely sensitive to uncertainties. The ability to reduce the cost reduction is severely hampered.  Although by $i=2$ the cost function appears smooth, the control-vector covariance has decreased significantly that all observations are ignored, indiscriminately, and the cost reduction remains approximately zero. 
In contrast, minimizing $\lambda_{\textrm{max}}$, and thus guarding against the largest curvature in $\hessian$, smooths the cost function, results in predictable convexity, and promotes convergence. 

\begin{figure}
    \centering
    \includegraphics[width=0.99\textwidth]{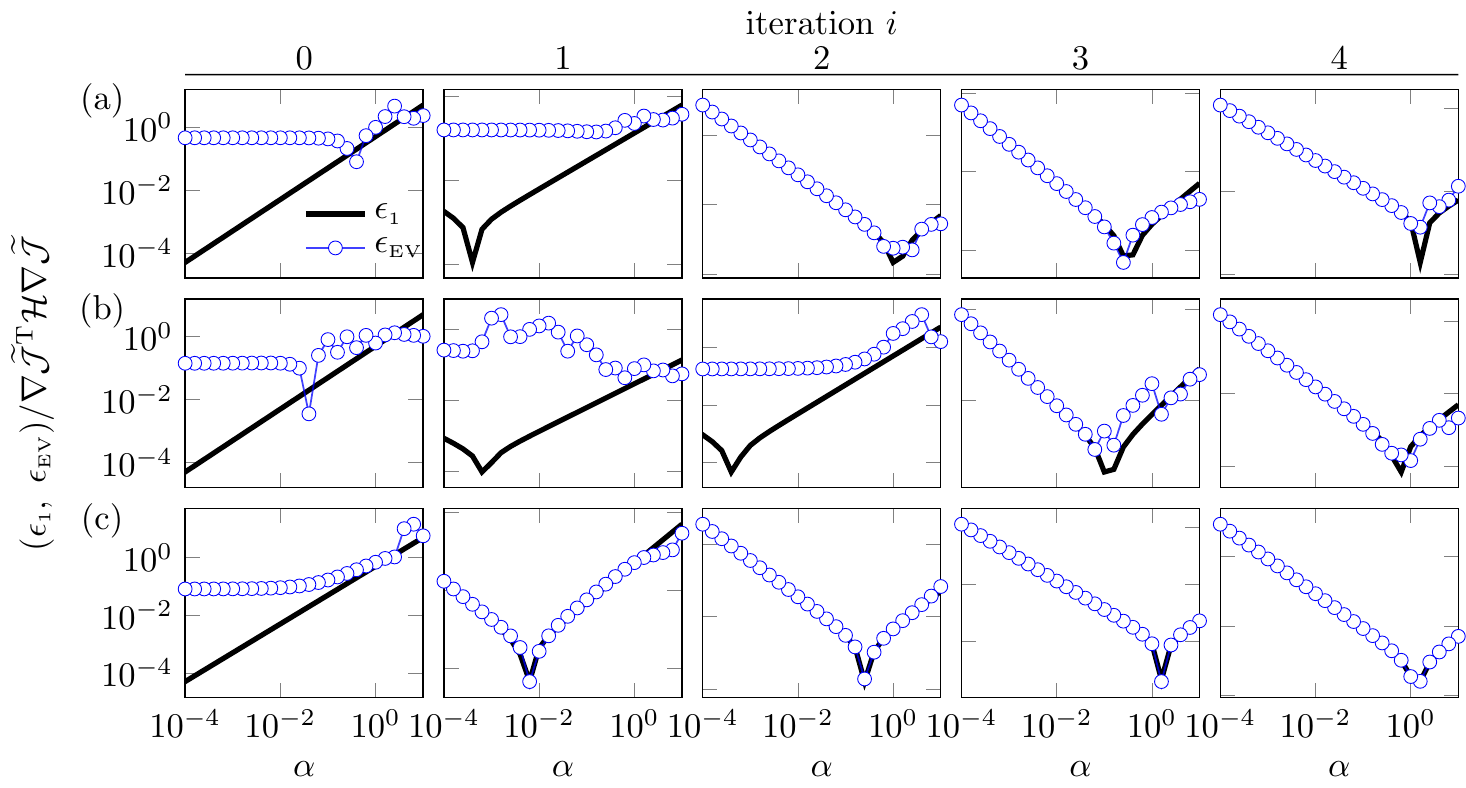}
    \caption{Accuracy of the gradient of the cost function evaluated using EnVar versus the step size in the optimal direction.  
    Results are shown for EnVar iterations $i\leq 4$. (a)~$\Weight=\identity$, (b)~$\Weight~|~\lambda_\mathrm{min}$, and (c)~$\Weight~|~\lambda_\mathrm{max}$.}
    \label{fig:lorenzGradientAccuracy}
\end{figure}

\begin{figure}
    \centering
    \includegraphics[width=0.99\textwidth]{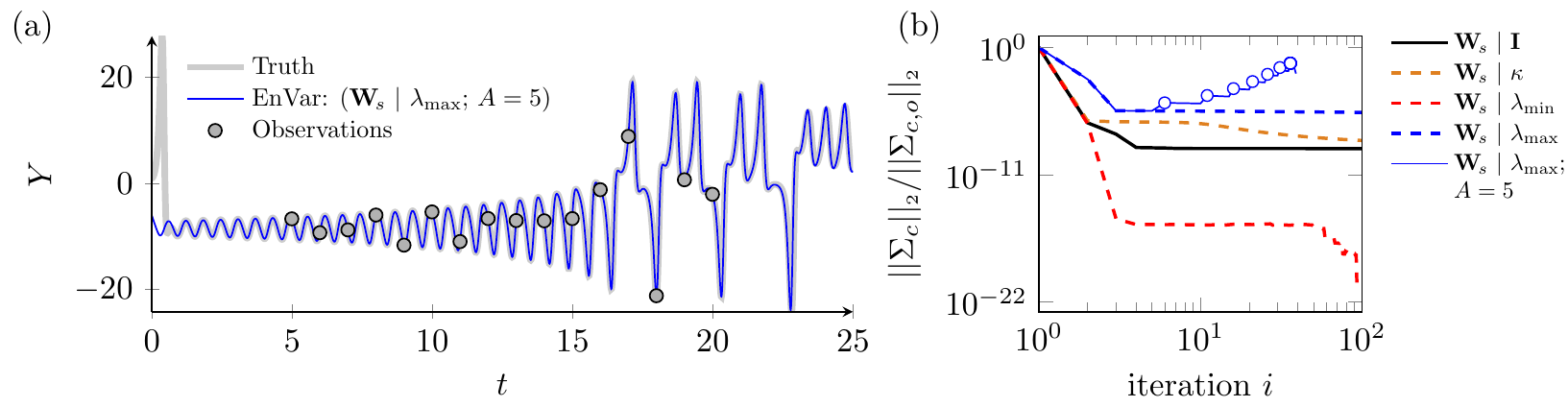}
    \caption{(a) The time evolution of $Y$ estimated using EnVar with weighted observations to guard against worst-case Hessian curvature, control-vector covariance inflation every five iterations, and stopping criteria $\cost/\cost_o \leq 10^{-11}$. (b) Normalized 2-norm of the control vector covariance matrix; symbols mark iterations when covariance inflation is applied.}
    \label{fig:covarianceMatrixLorenz}
\end{figure}

The degree to which nonlinearity affects gradient accuracy is quantified for the progression of EnVar iterations in figure~\ref{fig:lorenzGradientAccuracy}. We note that due to the initial sampling of $\covCntrVec=\sigma_c\identity$, the tangent approximation is not exactly obeyed, initially; though biasing earlier sensors in figure~\ref{fig:lorenzGradientAccuracy}~(c) shows improvement. After approximately 4 EnVar iterations, gradients are accurate and their accuracy is insensitive to weighting, which is due to decrease in $||\covCntrVec||$ (see figure~\ref{fig:covarianceMatrixLorenz}~b) and hence a refined local tangent approximation. Even with improved gradient accuracy by $i=4$, chaotic systems with oscillatory, or wrinkled, cost functions have local minima with large curvatures that tend to over-amplify trust in unknown control-vector parameters; the optimization thus stagnates at $\cost/\cost_o \approx 1$ (c.f.\, figure~\ref{fig:detailedCostLorenz}~(b,c)) because it is trapped in a local minimum. In contrast, 

We summarize by showing the final prediction of $Y$-dynamics in figure~\ref{fig:covarianceMatrixLorenz}~(a); trends for $X$ and $Z$ are similar, and are therefore not shown.  The predictions in the figure were obtained using the weighting scheme which minimizes $\lambda_{\mathrm{max}}$, or extreme curvature of the cost functional, and occasional inflation of the parameters covariance. The prediction matches the truth beyond the time horizon of the observations to $t=25$. \cite{Lorenz1963} notes that incomplete and imprecise data acquisition does not guarantee data match for $t\rightarrow\infty$; yet performance is surely enhanced. Notably, the optimized initial condition does not equal the truth, rather $\mathbf{X}_o$ was adjusted in a manner to reproduce the $\mathbf{X}$ dynamics by $t\approx 1$, which is earlier than the first observation at $t=5$.  Similar behavior was observed in boundary-layer transition in figure~\ref{fig:spectraInflow}~(e), which highlights the possibility of identifying local optima relative to the truth. Figure~\ref{fig:covarianceMatrixLorenz}~(b) shows how the recommended weighting avoids over-shrinking of $\covCntrVec$.  For uniform weighting, the parameter covariance matrix $\covCntrVec$ shrinks by approximately eight orders of magnitude by the third EnVar iteration figure~\ref{fig:detailedCostLorenz}~(b), due to sensitivity of observations to small changes in $\cntrVec$.  Without resorting \textit{ad hoc} methods of discarding observation data, weighting based on reducing extreme curvature favours earlier sensors and enhances the outcome of the optimization. 


\bibliography{./references}
\bibliographystyle{jfm}

\end{document}